\documentclass[apj]{emulateapj}

\shorttitle{The Cetus dSph galaxy}
\shortauthors{Monelli et al.}

\begin{document}

\title{The ACS LCID project. III. The star formation history of the 
Cetus dSph galaxy: \\ a post-reionization fossil.\altaffilmark{1}}

\author{M. Monelli\altaffilmark{2,3},
    S.L Hidalgo\altaffilmark{2,3},
    P.B. Stetson\altaffilmark{4},
    A. Aparicio\altaffilmark{2,3},
    C. Gallart\altaffilmark{2,3},
    A. Dolphin\altaffilmark{5},
    A. Cole\altaffilmark{6},
    D. Weisz\altaffilmark{7},
    E. Skillman\altaffilmark{7},
    E. Bernard\altaffilmark{8},
    L. Mayer\altaffilmark{9,10},
    J. Navarro\altaffilmark{11},
    S. Cassisi\altaffilmark{12},
    I. Drozdovsky\altaffilmark{2,3,13},
    E. Tolstoy\altaffilmark{14}
    }
    
\altaffiltext{1}{Based on observations made with the NASA/ESA {\it Hubble Space
   Telescope}, obtained at the Space Telescope Science Institute, which is
    operated by the Association of Universities for Research in Astronomy,
    Inc., under NASA contract NAS5-26555. These observations are associated
    with program 10505.}
\altaffiltext{2}{Instituto de Astrof\'{i}sica de Canarias, La Laguna, Tenerife,
    Spain; monelli@iac.es, carme@iac.es, dio@iac.es, antapaj@iac.es, slhidalgo@iac.es}
\altaffiltext{3}{Departamento de Astrof\'{i}sica, Universidad de La Laguna, 
    Tenerife, Spain}
\altaffiltext{4}{Dominion Astrophysical Observatory, Herzberg Institute of
    Astrophysics, National Research Council, Victoria, Canada;
    peter.stetson@nrc-cnrc.gc.ca}
\altaffiltext{5}{Raytheon; 1151 E. Hermans Rd., Tucson, AZ 85706, USA}
\altaffiltext{6}{School of Mathematics \& Physics, University of Tasmania,
    Hobart, Tasmania, Australia; andrew.cole@utas.edu.au}
\altaffiltext{7}{Department of Astronomy, University of Minnesota,
    Minneapolis, USA; skillman@astro.umn.edu}
\altaffiltext{8}{Institute for Astronomy, University of Edinburgh, Royal 
    Observatory, Blackford Hill, Edinburgh EH9 3HJ, UK; ejb@roe.ac.uk}
\altaffiltext{9}{Institut f\"ur Theoretische Physik, University of Zurich,
    Z\"urich, Switzerland; lucio@physik.unizh.ch}
\altaffiltext{10}{Department of Physics, Institut f\"ur Astronomie,
    ETH Z\"urich, Z\"urich, Switzerland; lucio@phys.ethz.ch}
\altaffiltext{11}{Department of Physics and Astronomy, University of 
    Victoria, 3800 Finnerty Road, Victoria, British Columbia, Canada V8P 5C2
    peter.stetson@nrc-cnrc.gc.ca}
\altaffiltext{12}{INAF-Osservatorio Astronomico di Collurania,
    Teramo, Italy; cassisi@oa-teramo.inaf.it}
\altaffiltext{13}{Astronomical Institute, St. Petersburg State University,
    St. Petersburg, Russia}
\altaffiltext{14}{Kapteyn Astronomical Institute, University of Groningen,
    Groningen, Netherlands; etolstoy@astro.rug.nl}

\begin{abstract}

We use deep HST/ACS observations to calculate the star formation history 
(SFH) of the Cetus dwarf spheroidal (dSph) galaxy. Our photometry reaches 
below the oldest main sequence turn-offs, which allows us to estimate 
the age and duration of the main episode of star formation in Cetus.
This is well approximated by a single episode that peaked roughly 12$\pm$0.5
Gyr ago and lasted no longer than about 1.9$\pm$0.5 Gyr (FWHM). Our 
solution also suggests that essentially no stars formed in Cetus during the
past 8 Gyrs. This makes Cetus' SFH comparable to that 
of the oldest Milky Way dSphs. Given the current isolation of Cetus in the outer 
fringes of the Local Group, this implies that Cetus is a clear outlier 
in the morphology-Galactocentric distance relation that holds for the 
majority of Milky Way dwarf satellites. 
%We find evidence for population 
%gradients within the field of view of our ACS observations, consistent 
%with a trend where star formation terminated earlier in the outer 
%regions of the dwarf. 
Our results also show that Cetus continued forming 
stars through $z \simeq$ 1, long after the Universe was reionized, and that 
there is no clear signature of the epoch of reionization in Cetus' SFH. 
We discuss briefly the implications of these results for dwarf galaxy 
evolution models. Finally, we present a comprehensive account of the 
data reduction and analysis strategy adopted for all galaxies targeted 
by the LCID (Local Cosmology from Isolated Dwarfs\footnotemark[15]) project. 
We employ two different photometry codes (DAOPHOT/ALLFRAME 
and DOLPHOT), three different SFH reconstruction codes (IAC-pop/MinnIAC,
MATCH, COLE), and two stellar evolution libraries (BaSTI and Padova/Girardi),
allowing for a detailed assessment of the modeling and observational 
uncertainties.

\footnotetext[15]{\itshape http://www.iac.es/project/LCID}

\end{abstract}


\section{Introduction}\label{sec:intro}

A powerful method to study the mechanisms that drive the evolution of
stellar systems is the recovery of their full star formation history (SFH). 
This can be done by coupling deep and accurate photometry, reaching the oldest 
main sequence (MS) turn-off (TO), with the synthetic color-magnitude diagram 
(CMD) modeling technique \citep{tosi91, bertelli92, tolstoy96, aparicio97, 
harris01, dolphin02, iacstar, cole07, iacpop}. The details of the first stages of galaxy 
evolution are particularly interesting, because they directly connect stellar 
populations research with cosmological studies. For example, the standard 
paradigm predicts that the role of re-ionization can impact the star formation 
history of small systems in a measurable way \citep[e.g.,][]{i86, r86, e92, br92,
cn94, qke96, tw96, kbs97, barkana99, bullock00, tassis03,
ricotti05, gnedin06, okamoto09}.

In this context, we designed a project aimed at recovering the full SFHs of 
six isolated galaxies in the Local Group (LG), namely IC~1613, Leo~A, LGS~3,
Phoenix, Cetus, and Tucana, with particular focus on the 
details of the early SFH. Isolated systems were selected because they are 
thought to have completed few, if any, orbits inside the LG. Therefore,
compared to satellite dSphs, they
spent most of their lifetime unperturbed, so that their evolution is 
expected not to be strongly affected by giant galaxies. 
A general summary of the goals, design and outcome of the LCID project can 
be found in Gallart et al. (in prep),  where a comparative study of the 
results for the six galaxies is presented. 

In this paper we focus on the stellar populations of the Cetus dSph galaxy only. 
This galaxy was discovered by \citet{whiting99}, by visual inspection of the 
ESO/SRC plates. From follow-up CCD observations they obtained a CMD to a 
depth of $V \sim 23$, clearly showing the upper part of the red giant branch (RGB), 
but no evidence of a young main sequence, nor of either red or blue supergiants 
typical of evolved young populations. From the tip of the RGB (TRGB) they could 
estimate a distance modulus $(m-M)_0 = 24.45 \pm 0.15$ mag ($776 \pm 53$ kpc),
assuming $E(B-V) = 0.03$ from the \citet{schlegel98} maps and an absolute 
magnitude $M_{I}(TRGB) = -3.98 \pm 0.05$ mag for the tip. The color of the
RGB stars near the tip allowed a mean metallicity estimate of $[Fe/H] = -1.7$
dex with a spread of $\sim 0.2$ dex. Deeper HST/WFPC2 data presented by 
\citet{sarajedini02} confirmed the distance modulus of Cetus: 
$(m-M)_0 = 24.49 \pm 0.14$ mag, or $790 \pm 50$ kpc, assuming $M_{I}(TRGB) 
= -4.05 \pm 0.10$ mag. Based on the color distribution of bright RGB stars, 
they estimated a mean metallicity $[Fe/H] = -1.9$ dex. An interesting 
feature of the Cetus CMD presented by \citet{sarajedini02} is that the 
horizontal branch (HB) seems to be more populated in the red than in the blue part 
\citep[see also the CMD by][based on VLT/FORS1 data]{tolstoycetus00}. 
\citet{sarajedini02} calculated an HB $index = -0.91 \pm 0.09$\footnotemark[16].
By using an empirical relation between 
the mean color of the HB and its morphological type, derived from globular 
clusters of metallicity similar to that of Cetus, they concluded that the Cetus 
HB is too red for its metallicity, and therefore, is to some extent affected by 
the so-called second parameter. Assuming that this is mostly due to age, they 
proposed that Cetus could be 2-3 Gyr younger than the old Galactic 
globular clusters.

\footnotetext[16]{The HB index is defined as ($B$-$R$)/($B$+$V$+$R$), \citep{lee90},
where B and R are the number of HB stars bluer and redder than the instability 
strip, and V is the total number of RR Lyrae stars.}

Recent work by \citet{lewis07} presented wide-field photometry from the 
Isaac Newton Telescope together with Keck spectroscopic data for the 
brightest RGB stars. From CaT measurements of 70 stars, they estimated 
a mean metallicity of $[Fe/H] = -1.9$ dex, fully consistent with previous estimates 
based on photometric data. Moreover, they determined for the first time a systemic 
heliocentric velocity of $\sim$ 87 km~s$^{-1}$, an internal velocity dispersion 
of 17 km~s$^{-1}$ and, interestingly enough, little evidence of rotation.

 The search for H~I gas \citep{bouchard06} revelead three possible candidate
clouds associated with Cetus, located beyond its tidal radius, but within 20 Kpc
from its centre.
Considering the velocity of these clouds, \citet{bouchard06} suggested that
the velocity of Cetus should be of the order of -280$\pm 40 km~s^{-1}$ to
have a high chance of association. Clearly, this does not agree with the estimates 
later given by \citet{lewis07}, who pointed out that Cetus is devoid of gas, at 
least to the actual observational limits.

Cetus is particularly interesting for two distinctive characteristics. 
First, the high degree of isolation, it being at least $\sim 680$ kpc away 
from both the Galaxy and M31. Cetus is, together with Tucana, one of 
the two isolated dwarf spheroidals in the LG. Moreover, based on 
wide-field INT telescope data, \citet{mcconnachie06} estimated a tidal radius of 
6.6 kpc, which would make Cetus the largest dSph in the LG.
Therefore, an in-depth study of its stellar populations, and how they 
evolved with time, could provide important clues about the processes 
governing dwarf galaxy evolution.

The paper is organized as follows. In \S \ref{sec:obse} we present 
the data and the reduction strategy. \S \ref{sec:cmd} discusses the
details of the derived color-magnitude diagram (CMD). In \S 
\ref{sec:methods} we summarize the methods adopted to recover the SFH,
and we discuss the effect of the photometry/library/code on the
derived SFH. The Cetus SFH is presented in \S \ref{sec:results},
where we discuss the details of the results, as well as the tests
we performed to estimate the duration of the main peak of star formation and the
radial gradients. A discussion of these results in the general context
of galaxy evolution are presented in \S \ref{sec:discu}, and
our conclusions are summarized in \S \ref{sec:summa}.
 
\section{Observations and data reduction strategy}\label{sec:obse}

The data presented in this paper were collected with the ACS/WFC
camera \citep{ford98} aboard the HST. The observations were designed 
to obtain a signal-to-noise ratio $\approx 10$ at the magnitude level 
of the oldest MSTO, $M_{F814W} \sim +3$. Following the precepts 
outlined in \citet{stetson93}, the choice of the filters was based 
on the analysis of synthetic CMDs in the ACS bands. The $(F475W-F814W)$ 
pair, due to the large color baseline together with the large filter 
widths, turned out to be optimal for separating age differences 
of the order of 1-2 Gyr in stellar populations older than 10 Gyr.

The observations were split into two-orbit visits. One $F475W$ 
and one $F814W$ image were collected during each orbit, with exposure 
times slightly different between the two orbits of the same visit:
1,280 s and 1,135 s, respectively, in the first orbit, 1,300 s and 1,137 s 
during the second one. The last visit included a third orbit, with exposure 
times of 1,300 s and 1,137 s. The 27 orbits allocated to this galaxy were 
executed between August 28 and 30, 2006. The two orbits corresponding to 
the ninth visit (images j9fz27glq, j9fz27gmq, j9fz27goq, j9fz27gqq) 
suffered from loss of the guide star. Therefore, two images were 
found to be shifted by $\approx$1.5$\arcmin$, and the other two were aborted after
a few seconds. These data will not be included in the following analysis.
Summarizing, the total exposure time devoted to Cetus and used in this work was
32,280 s in $F475W$ and 28,381 s in $F814W$.

\begin{figure}
\epsscale{.80}
%\plotone{map.eps}
\plotone{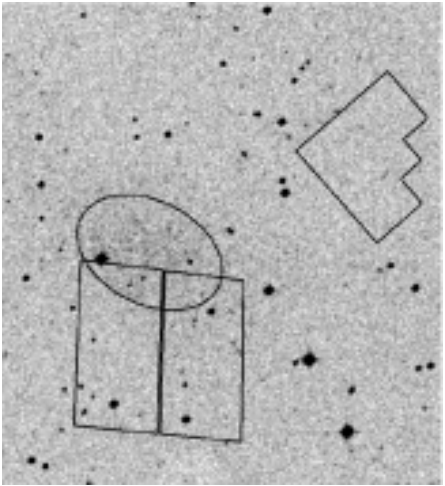}
\caption{9$\arcmin \times$ 9$\arcmin$ DSS image showing the ACS and the WFPC2 
pointings. The ellipse represents the Cetus core radius. \label{fig:map}}
\end{figure}

\begin{figure*}
\includegraphics[scale=.80]{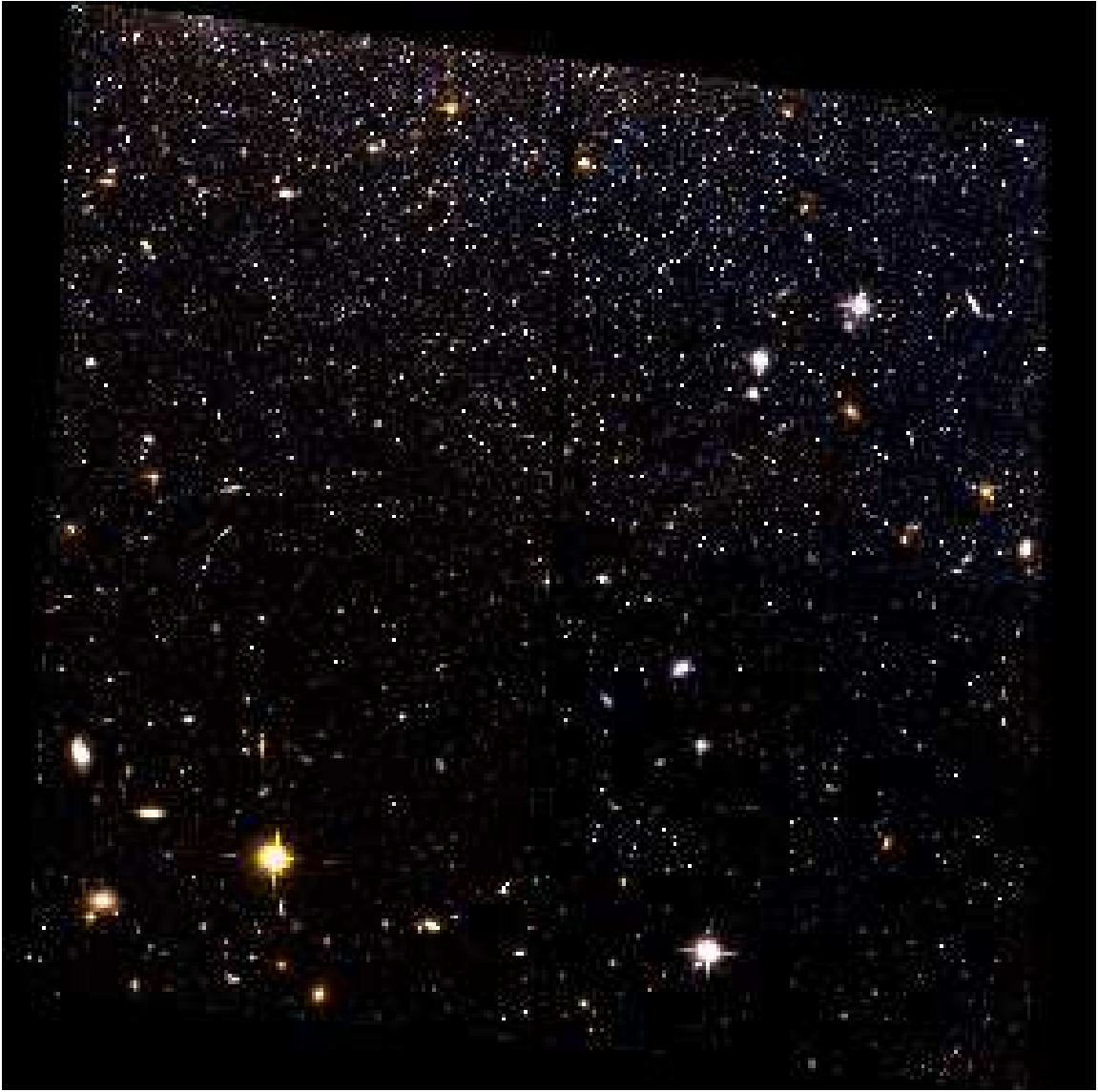}
\caption{Stacked, drizzled  color image of the Cetus field. North is up 
and East is left.  The field of view is $\sim$3.4$\times$3.4 square arcmin. 
The image shows a clear gradient in the number of stars when moving toward the 
south. Note the considerable number of background galaxies.
\label{fig:drz}}
\end{figure*}

For our observing strategy, we split each orbit into one $F475W$ and one $F814W$ 
exposure in order to have
the best possible time sampling for short-period variable stars
\citep[see][]{bernardcetustucana}. An off-center pointing was chosen to 
sample a wide range of galactocentric radii. The edge of the ACS camera was 
shifted to the south by $\sim2\arcmin$ from the center of Cetus, in order 
to avoid a bright field star. The coordinates of the ACS pointings are 
[$RA = 00^h 26^m 09^s.70, DEC= -11\arcdeg 04\arcmin 34.7\arcsec $]. Fig. 
\ref{fig:map} shows the location of both the ACS and the parallel WFPC2 field
[$RA = 00^h 25^m 51^s.41, DEC = -11^\circ 00\arcmin 32.4\arcsec$]. The pointing 
of the latter was optimized to explore the outskirts of the galaxy, and the 
resulting CMD has already been presented in \citet{bernardcetustucana}.
Due to the much shallower photometry in the WFPC2 field, this will not be 
included in the analysis of the SFH. Fig. \ref{fig:drz} shows a color drizzled,
stacked image. Note the strong gradient in the number of stars
when moving outward toward the south. There are also a few bright field 
stars, and a sizable number of extended background sources.

%%%%%%%%%%%%%%%%%%%%%%%%%%%%%%%%%%%%%%%%%%%%%%%%%%%%%%%%%%%%%%%%%%%%%%
%%%%%%%%%%%%%%%%%%%%%%%%%%%%% FIG 1 %%%%%%%%%%%%%%%%%%%%%%%%%%%%%%%%%%

%%%%%%%%%%%%%%%%%%%%%%%%%%%%%%%%%%%%%%%%%%%%%%%%%%%%%%%%%%%%%%%%%%%%%%
%%%%%%%%%%%%%%%%%%%%%%%%%%%%%%%%%%%%%%%%%%%%%%%%%%%%%%%%%%%%%%%%%%%%%%

%%%%%%%%%%%%%%%%%%%%%%%%%%%%%%%%%%%%%%%%%%%%%%%%%%%%%%%%%%%%%%%%%%%%%%
%%%%%%%%%%%%%%%%%%%%%%%%%%%%% FIG 2 %%%%%%%%%%%%%%%%%%%%%%%%%%%%%%%%%%

%%%%%%%%%%%%%%%%%%%%%%%%%%%%%%%%%%%%%%%%%%%%%%%%%%%%%%%%%%%%%%%%%%%%%%
%%%%%%%%%%%%%%%%%%%%%%%%%%%%%%%%%%%%%%%%%%%%%%%%%%%%%%%%%%%%%%%%%%%%%%

In the following, we outline the homogeneous reduction strategy adopted 
for all the galaxies of the LCID project. In particular, we stress that we 
performed two parallel and independent photometric reductions, based on 
the DAOPHOT/ALLFRAME  \citep{alf} and the DOLPHOT packages \citep{dolphot},
in order to investigate the impact of using different photometry codes to 
obtain the CMD used to derive the SFH (see \S \ref{sec:sfh_comparison}).

	\subsection{DAOPHOT reduction}

After testing different approaches to drizzle the set of images,
we opted to work on the original \_FLT data.  A custom dithering  
pattern was adopted for the observations. The shifts applied were
small enough ($<10$ pixels) that the camera distortions 
would not cause problems when 
subsequently merging the catalogs from different images. The two 
chips of the camera have been treated independently, and we applied 
the following steps to all the individual images:

$\bullet$ {\itshape Basic reduction:} we adopted the basic reduction 
(bias, flat field) from the On-The-Fly-Reprocessing pipeline (CALACS 
v.4.6.1, March 13 2006). Since we opted for the \_FLT images, we 
applied the pixel area mask to account for the different  area of the 
sky projected onto each pixel, depending on its position;

$\bullet$ {\itshape Cosmic ray rejection:} we used the data quality 
files from the pipeline to mask the cosmic rays (CR). According to 
the flag definition in the ACS Data handbook, pixels with the value 4096 
are affected by cosmic rays. These pixels were flagged with a value 
higher than the saturation threshold. This allowed DAOPHOT to recognize 
and properly ignore them;

$\bullet$ {\itshape Aperture photometry:} we performed the source 
detection at a 3-$\sigma$ level and aperture photometry, with aperture radius=3 pixels;

$\bullet$ {\itshape PSF modeling:} careful selection of the PSF 
stars was done through an iterative procedure. Automated routines were 
used to identify bright, isolated stars, with good shape parameters 
({\itshape sharpness}) and small photometric errors. PSF stars were 
selected in order to sample the whole area of the chip, to properly take
into account the spatial variations of the PSF. The last step was a 
visual inspection of all the PSF stars, for all the images, to reject 
those with peculiarities like bright neighbors,  
contamination by CRs, or damaged columns. On average, we ended up with 
more than 150 PSF stars per image.

We performed many tests in order to understand which PSF model, 
among those available in DAOPHOT, would work best with our data. 
We used the same sets of stars for each fixed analytical PSF model, 
and tested different degrees of variation across the field, from 
purely analytical and constant ($variability=-1, 0$) to varying cubically ($va=3$).
We evaluated the quality of the final CMD, and the best results were 
obtained with $va>0$, but the tightness of the CMD sequences did not vary 
appreciably when using a quadratically or cubically variable PSF. We 
also checked the photometric errors and sharpness estimators, but 
could not find any evident improvement when increasing the degree of 
variation of the PSF. We therefore adopted a simpler and faster 
linear PSF, allowing the software to choose the analytical function 
on the basis of the $\chi^2$ estimated by the $PSF$ routine. In all 
the images, a Moffat function with index $\beta$ = 1.5 was selected.

The next step was ALLSTAR profile-fitting photometry for individual 
images. The resulting catalogs were used only to calculate the 
geometric transformations with respect to a common coordinate system, 
using DAOMATCH and DAOMASTER. Geometrical transformations are 
needed to generate a stacked image with MONTAGE2. The resulting median 
CR-cleaned image was used to create the input list for ALLFRAME. 
Rejecting all the objects with $|sharp| > 0.3$ was an efficient 
way to remove most of the extended objects (i.e., background galaxies), 
as well as remaining cosmic rays. 
After the first ALLFRAME was run, we used the updated and more precise 
positions and magnitudes to refine the coordinate transformations and 
to recalculate the PSFs, using the previously selected list of stars. 
A second ALLFRAME run was finally ran with updated geometric transformations
and PSFs.

The calibration to the VEGAMAG system was done following the prescriptions 
given by \citet{sirianni05}, using the updated zero points from \cite{mack07} (because
Cetus was observed after the temperature change of the camera in July 2006). 
Particular attention was paid to calculate the aperture correction
to the default aperture radius of 0.5\arcsec. Aperture photometry in various apertures
was performed on the PSF stars of every image, once the neighboring stars had
been subtracted. The growth curve and total magnitude of the stars were obtained
using DAOGROW \citep{daogrow}. An aperture correction was calculated for every image, 
with the result that the mean correction was 0.029 $\pm$ 0.009 mag.
Individual catalogs were calibrated, and the final list of stars was obtained 
keeping all the objects present in at least {\itshape (half+1)} images in both 
bands. The final CMD is presented and discussed in \S \ref{sec:cmd}.

	\subsection{Crowding tests with DAOPHOT}\label{sec:crow_dao}

To estimate the completeness and errors affecting our photometry, we 
adopted a standard method of artificial-star tests. We added 
$\sim$ 350,000 stars per chip, adopting a regular grid where the stars were 
located at the vertices of equilateral triangles. The distance between 
two artificial stars was fixed in order to pack the largest number of 
stars but at the same time to ensure that the synthetic stars were not 
affecting each other. This value was chosen to be {\itshape R =  
($rad_{PSF}$ + $rad_{fit}$ + 1)} = 13 pixels. In this way, overlapping of 
the wings of the artificial stars was minimized to avoid influencing  
the fit in the core of the PSF.
With this criterion, taking into account the dimension of the CCD, 
we needed seven iterations with $\sim$ 49,770 stars each. In each 
iteration, the grid was moved by a few pixels, to better sample the 
crowding characteristics of the image.

To verify the hypothesis that the synthetic stars were not affecting 
each other, we also performed completeness tests injecting stars separated 
by {\itshape ($2 \times rad_{PSF}$ + 1)} = 21 pixels. We did not find any 
significant difference in the completeness level of synthetic stars,
meaning that the original separation
was adequate. With this criterion, the incompleteness of the synthetic 
stars only depends on the real stars, while the additional crowding
due to the partial overlap of the injected stars is negligible.

\begin{figure}
\epsscale{1.5}
%\plottwo{demag3a.eps}{demag4a.eps}
\plottwo{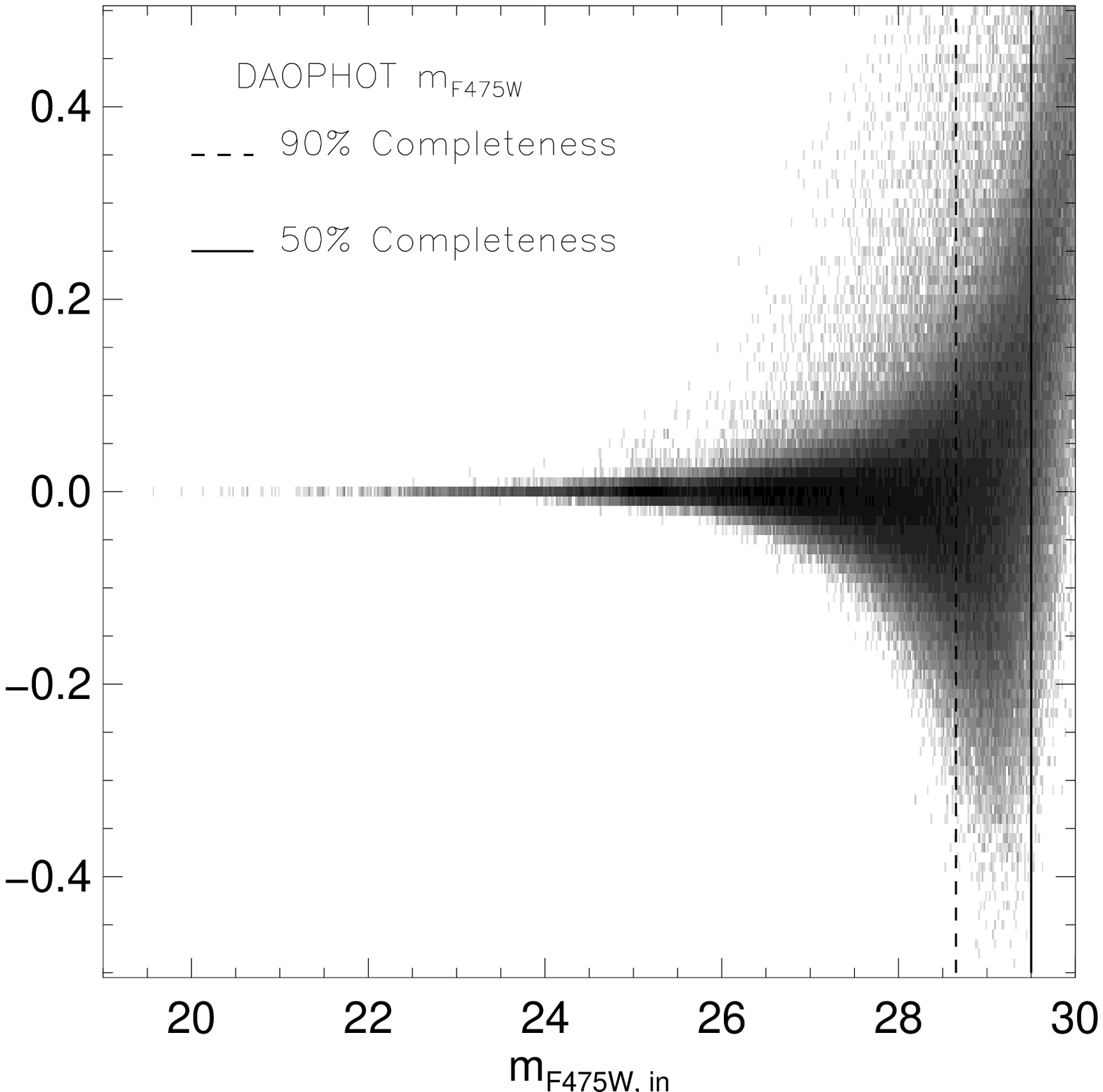}{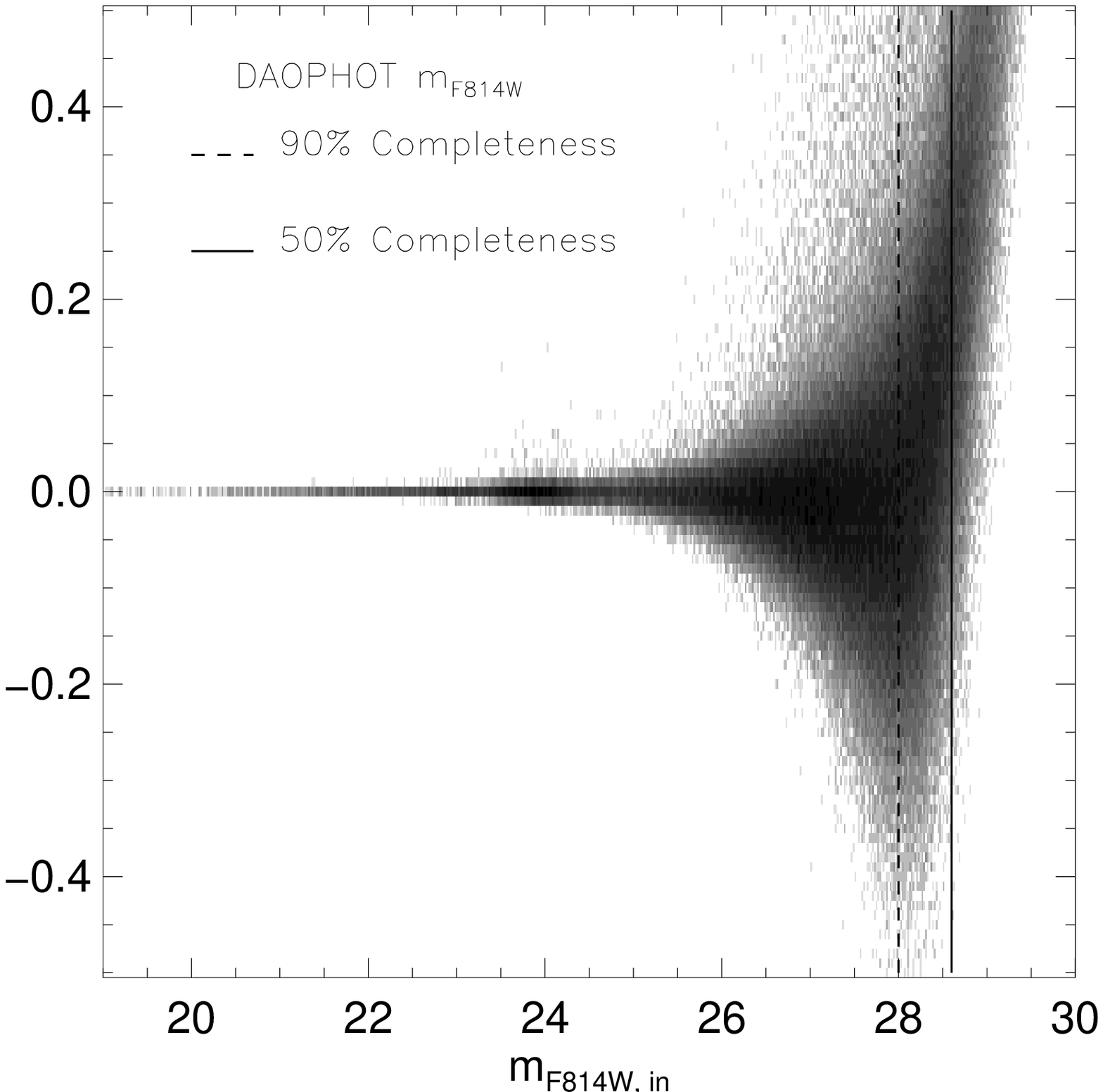}
\caption{Difference $\Delta mag$=mag$_{in}$-mag$_{out}$ for synthetic 
stars. The two vertical lines mark the 90\% and 50\% completeness level.
A total of $\sim$700,000 stars were simulated.
\label{fig:demag}}
\end{figure}

The list of injected stars was selected following the prescriptions 
introduced by \citet{gallart96}. To fully sample the observed magnitude 
and color ranges, we adopted a synthetic CMD created with IAC-star 
\citep{iacstar}, adopting a constant star formation rate (SFR) between 0 
and 15 Gyr, with the metallicity uniformly distributed between Z=0.0001 
and Z=0.005 at any age. The stars were added on the individual 
images, taking into account the coordinate and magnitude shifts.
The photometry was then repeated with the same prescriptions as for 
the original images. The results are summarized in Fig. \ref{fig:demag},
where the $\Delta mag$=mag$_{in}$-mag$_{out}$ of the artificial 
stars and selected completeness levels are shown for the two bands, as
a function of the input magnitude.

%%%%%%%%%%%%%%%%%%%%%%%%%%%%%%%%%%%%%%%%%%%%%%%%%%%%%%%%%%%%%%%%%%%%%%
%%%%%%%%%%%%%%%%%%%%%%%%%%%%% FIG 3 %%%%%%%%%%%%%%%%%%%%%%%%%%%%%%%%%%

%%%%%%%%%%%%%%%%%%%%%%%%%%%%%%%%%%%%%%%%%%%%%%%%%%%%%%%%%%%%%%%%%%%%%%
%%%%%%%%%%%%%%%%%%%%%%%%%%%%%%%%%%%%%%%%%%%%%%%%%%%%%%%%%%%%%%%%%%%%%%

%%%%%%%%%%%%%%%%%%%%%%%%%%%%%%%%%%%%%%%%%%%%%%%%%%%%%%%%%%%%%%%%%%%%%%
%%%%%%%%%%%%%%%%%%%%%%%%%%%%% FIG 4 %%%%%%%%%%%%%%%%%%%%%%%%%%%%%%%%%%

\begin{figure}
\epsscale{1.5}
%\plottwo{demag3a_dolp.eps}{demag4a_dolp.eps}
\plottwo{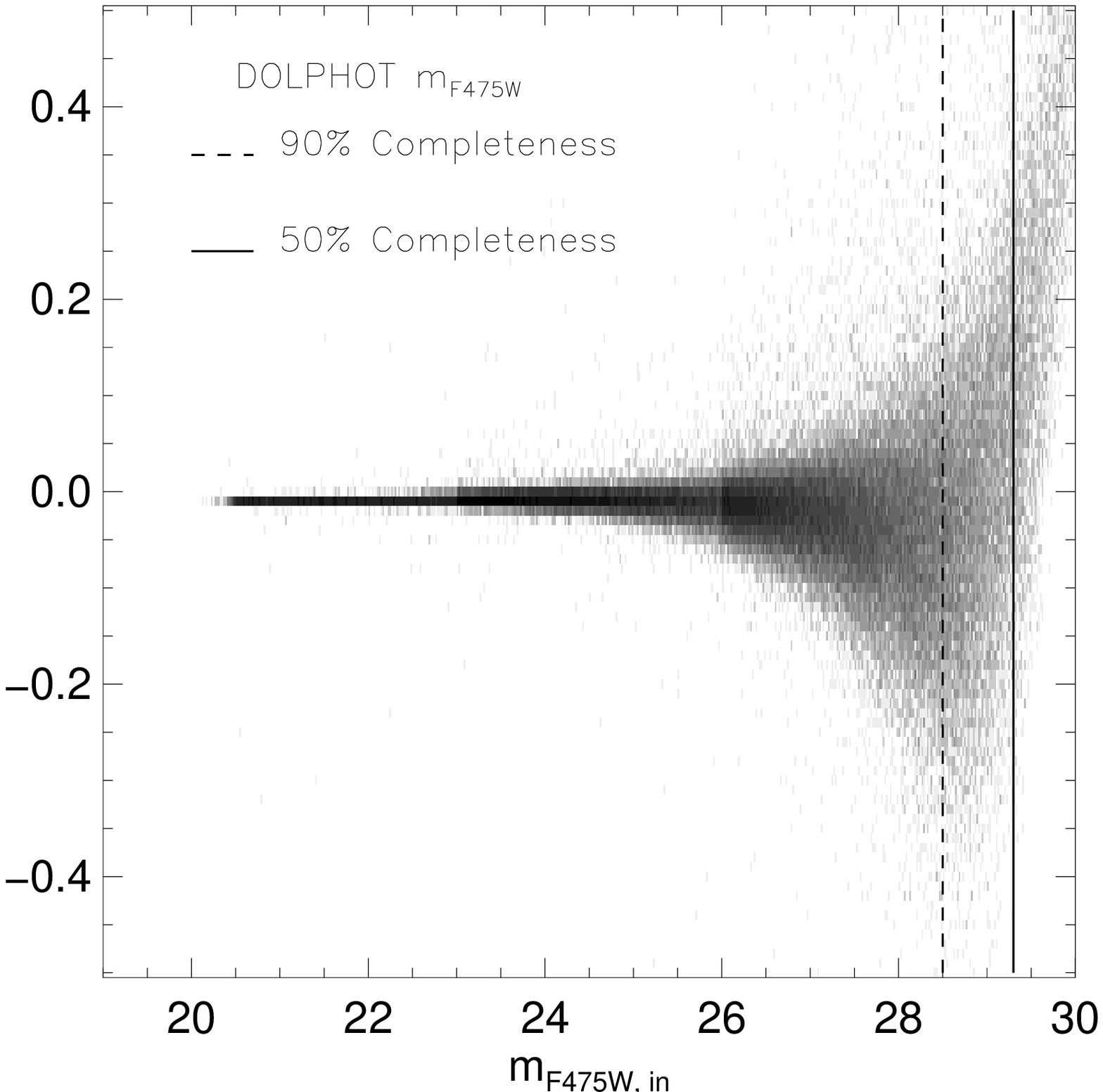}{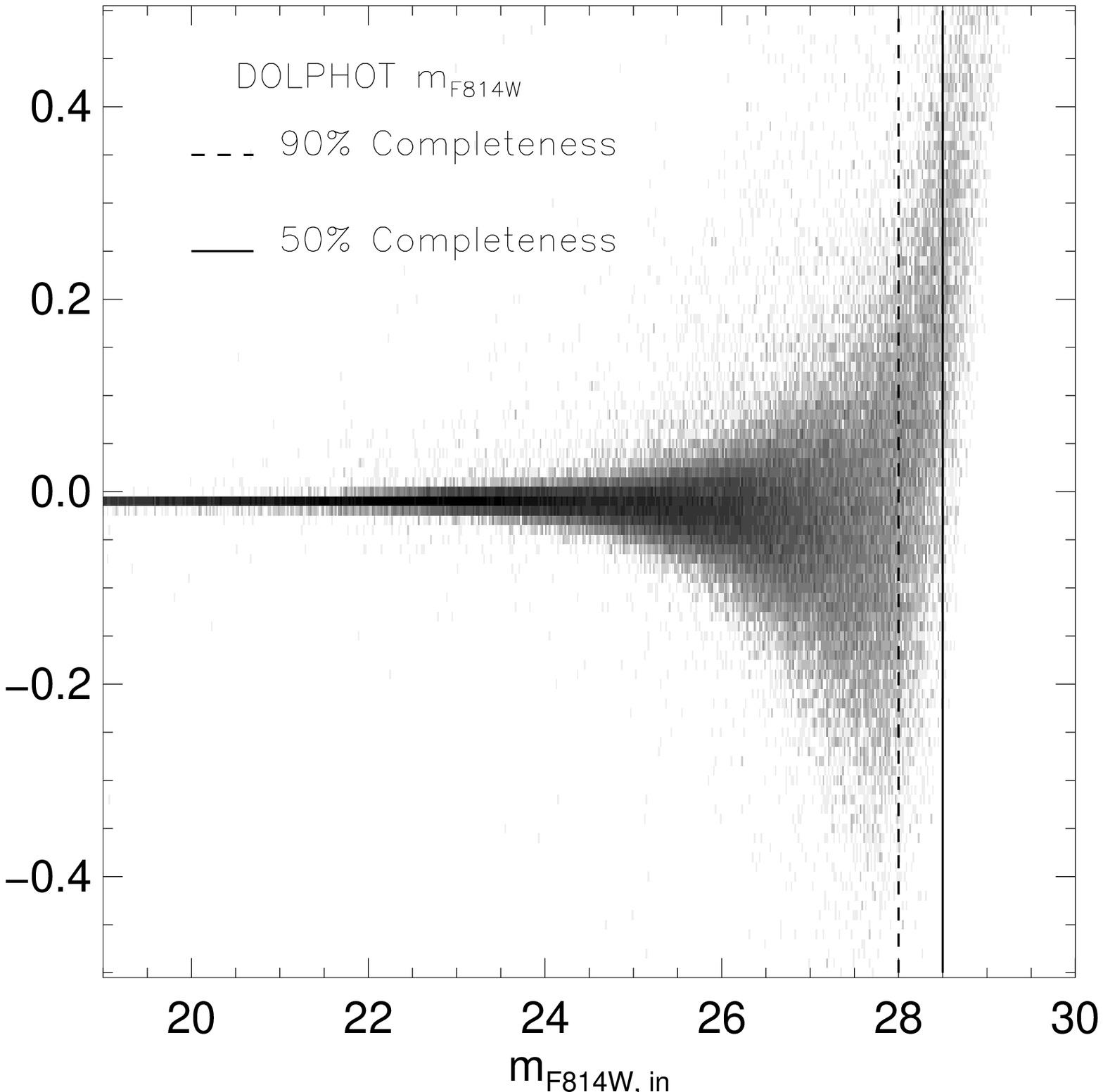}
\caption{Same as Fig. \ref{fig:demag}, but for the DOLPHOT photometry.
A total of $\sim$140,000 stars were simulated.
\label{fig:demag_dol}}
\end{figure}

%%%%%%%%%%%%%%%%%%%%%%%%%%%%%%%%%%%%%%%%%%%%%%%%%%%%%%%%%%%%%%%%%%%%%%
%%%%%%%%%%%%%%%%%%%%%%%%%%%%%%%%%%%%%%%%%%%%%%%%%%%%%%%%%%%%%%%%%%%%%%

	\subsection{DOLPHOT reduction}

DOLPHOT is a general-purpose photometry package adapted from the
WFPC2-specific package HSTphot \citep{dolphot}.\footnotemark[17]
For this reduction, we used the ACS module of DOLPHOT and followed the 
recommended photometry recipe provided in the manual for version 1.0.3.  
The procedure is fairly automated, with the software locating stars, 
adjusting the default Tiny Tim PSFs based on the image, iterating until 
the photometry converges, and making aperture corrections. 
Similar to the procedures with DAOPHOT, we used the 
original \_FLT images, corrected for the pixel-area mask
and cosmic rays. We used a drizzled image from the HST archive as the
reference for the coordinate transformations. The search for 
stellar sources was performed with a threshold $t$=2.5$\sigma$.
The photometry was done fitting a model PSF calculated with TinyTim.
DOLPHOT also automatically estimated the aperture correction to 
0.5$\arcsec$. The same calibration as for the DAOPHOT photometry was used.

\footnotetext[17]{The code is publicly available from 
{\itshape http://purcell.as.arizona.edu/dolphot/ }.}

	\subsection{Crowding tests with DOLPHOT}\label{sec:crow_dol}

Artificial-star tests were preformed by injecting $\sim 140,000$ synthetic 
stars in the original images. The input colors and magnitudes of 
the synthetic stars cover the complete range of the observed 
colors and magnitudes. The stars uniformly sample the range $-1 < M_{F475W} - 
M_{814W} < 5$ mag, $-7 < M_{F475W} < 8$ mag. The photometry was 
performed following the procedure suggested by \citet{holtzman06}. The 
synthetic stars were injected one at a time into the images to avoid 
affecting the stellar crowding in the images. Stars were uniformly 
distributed to cover the whole area of the camera. Fig. \ref{fig:demag_dol} 
shows $\Delta mag$=mag$_{in}$-mag$_{out}$ of the artificial 
stars together with the 50\% and 90\% completeness levels for the two bands.

	\subsection{Comparison of the two photometry sets}

\begin{figure}
\epsscale{1.2}
%\plotone{delta_cetus5.eps}
\plotone{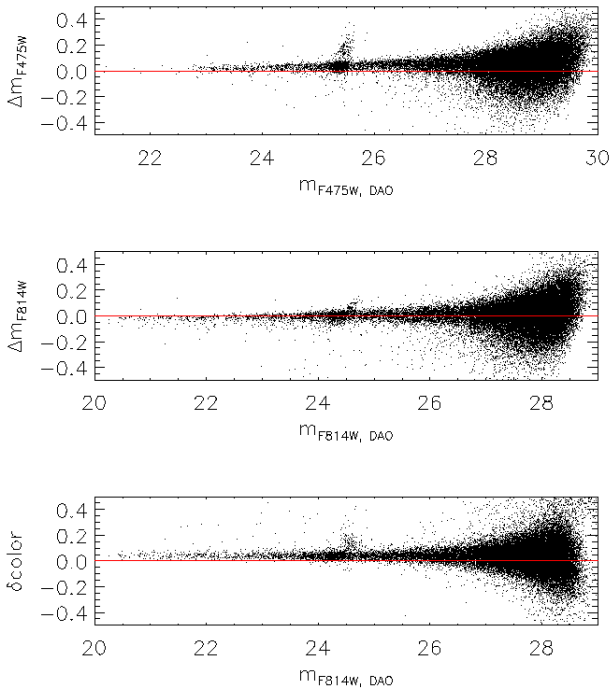}
\caption{Magnitude and color difference between the DAOPHOT and DOLPHOT 
photometry, as a function of the DAOPHOT magnitude. We detect a 
systematic residual zero point between the two codes, and a small 
trend as a function of magnitude. At the faintest end, DAOPHOT gives 
fainter measurements than DOLPHOT. However, since the trend is similar 
in the two filters, the color difference showed only a small residual 
zero point of the order of 0.04 mag. The plume of objects at 
$m_{F475W} \sim 25.5$ is made of RR Lyrae variable stars.
\label{fig:delta_cetus5}}
\end{figure}

A direct comparison between the DAOPHOT and DOLPHOT photometry is 
presented in Fig. \ref{fig:delta_cetus5}. We detect a small zero point offset
for the brightest stars in both bands ($(m_{F475W,DAO}-m_{F475W,DOL}) \sim 0.023$ 
mag, for $m_{F475W} <$ 25, and $(m_{F814W,DAO}-m_{F814W,DOL}) \sim -0.010$ mag, 
for $m_{F814W} <$ 24), and a trend as a function of magnitude: moving 
toward fainter magnitudes, DAOPHOT tends to give slightly fainter 
results (up to $\Delta m_{F475W}\simeq$ 0.05 mag, at $m_{F475W} = 29.1$,
$\Delta m_{F814W} \simeq 0.07$ at $m_{F814W} = 28.0$). Note that since the trend with 
magnitude is similar in the two bands, the color difference shows a 
residual zero point offset ($\sim0.04$ mag), but no dependence on magnitude.
We performed an exhaustive search for the possible cause of such
small differences, but its origin remained elusive, and it is likely 
a result of the many differences between the approaches of the two
photometric packages (PSF, sky determination, aperture correction calculation).
Note that similar trends were detected and discussed by \citet{holtzman06} 
and \citet{hill98} when comparing different photometric codes applied 
to HST data. This reinforces the idea that small differences in the 
resulting photometry naturally arise when using different reduction
codes, at least when dealing with HST data.

However, we highlight that, for this project, the MinnIAC/IAC-pop 
method presented in \ref{sec:method_iacpop} is not sensitive to 
zero-point systematics, and therefore the impact of
small differences in the photometry on the derived SFH is minimized.

%%%%%%%%%%%%%%%%%%%%%%%%%%%%%%%%%%%%%%%%%%%%%%%%%%%%%%%%%%%%%%%%%%%%%%
%%%%%%%%%%%%%%%%%%%%%%%%%%%%% FIG 5 %%%%%%%%%%%%%%%%%%%%%%%%%%%%%%%%%%

%%%%%%%%%%%%%%%%%%%%%%%%%%%%%%%%%%%%%%%%%%%%%%%%%%%%%%%%%%%%%%%%%%%%%%
%%%%%%%%%%%%%%%%%%%%%%%%%%%%%%%%%%%%%%%%%%%%%%%%%%%%%%%%%%%%%%%%%%%%%%

\section{The color-magnitude diagram} \label{sec:cmd}

\begin{figure*}[t]
\epsscale{1.0}
%\plotone{cetus_noiso_both_errcross.eps}
\plotone{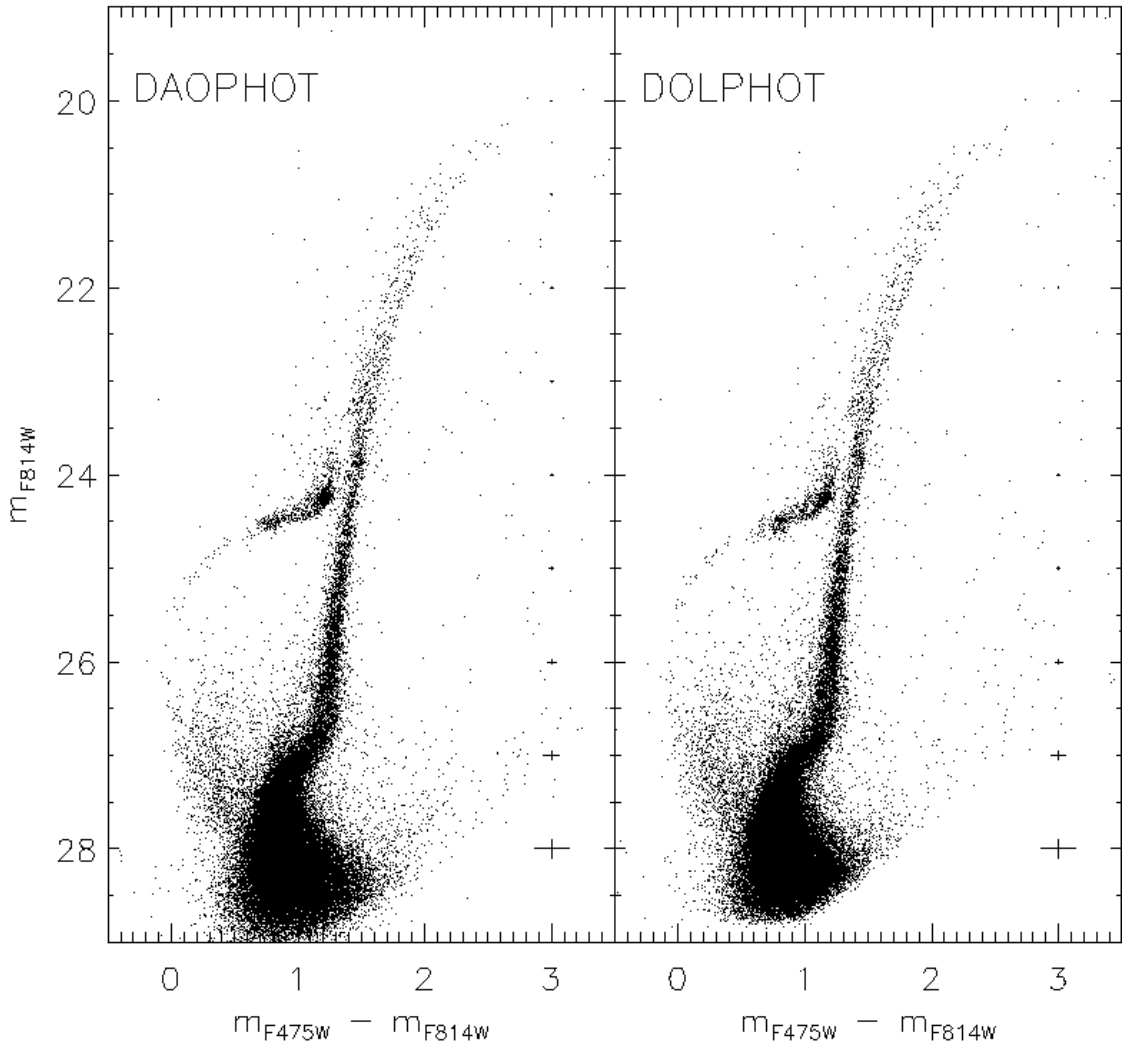}
\caption{Color-magnitude diagram, calibrated in the VEGAMAG photometric system,
for both the DAOPHOT and DOLPHOT photometry. On the right side of both panels 
we report the mean magnitude and color fitting errors provided by the two codes.
\label{fig:cmd_noiso}}
\end{figure*}

Fig.~\ref{fig:cmd_noiso} presents the final CMD from both the DAOPHOT 
and DOLPHOT photometry, calibrated to the VEGAMAG system. 
This is the deepest CMD obtained to date for this galaxy, spanning more 
than 8 mag, from the TRGB down to $\approx$1.5 mag below the oldest MSTO.
The limiting magnitude (F814W $\sim$ 28.8 mag) is similar in both 
diagrams. No selections were applied to the DAOPHOT CMD other than those already 
applied to the star list, while the DOLPHOT photometry was cleaned according to 
the  {\itshape sharpness} ($|sharp| < 0.1$), {\itshape crowding} ($crowd <0.7$),
and photometric errors ($\sigma_{m_{F475W}}, \sigma_{m_{F814W}} < 0.2$).
The DAOPHOT and DOLPHOT CMDs contain $\sim$ 50,600 and 44,900 stars, respectively

A simple qualitative analysis reveals many interesting features.
The TO emerges, for the first time and very well characterized in our photometry,
at $m_{F814W} \sim$ 27 mag. The morphologies of the TO and the sub-giant branch
region suggest that Cetus is a predominantly old system. 
There is no evidence of bright main sequence stars indicative of recent star 
formation. A plume of relatively 
faint objects appears at 26 $< m_{F814W} <$ 28, 0 $< (m_{F475W} - m_{F814W}) <$ 0.5, 
and could indicate either a low level of star formation until $\sim$ 2-3
Gyr ago, or the presence of a population of blue straggler stars (BSs)
\citep[cf.,][]{mapelli07, momany07, mapelli09}.
We noticed that the over-density right above the TO, at $0.8 < (m_{F475W} - m_{F814W}) 
< 1, 26.5 < m_{F814W} < 27.0$, is mostly present in the central
regions. We verified that, with the DAOPHOT photometry, such overdensity 
appears as well in the CMD of the synthetic stars used in the completenss tests
and located in the most crowded areas of the images.
Since we simulated the same CMD in all the regions of the camera, there is no reason
to expect differences in the output synthetic CMDs as a function of the position.
This indicates that the higher crowding in the central region is responsible for 
this feature. Nevertheless, it is possible that some of the stars located in this region are physical binaries.

Concerning the evolved phases of stellar evolution, the RGB between 20 
$< m_{F814W} <$ 26.5 mag appears as a prominent feature. The width of the 
RGB is similar in both diagrams, and suggests some spread
in age and/or metallicity.
The sudden increase in the RGB width at $m_{F814W} \leq $ 23.2 may be 
explained by the superposition of the RGB and the asymptotic giant branch
(observational effects like saturation of bright stars or loss of linearity of the camera 
can be safely excluded in this regime). 

The HB, as already noted by \citet{sarajedini02} and \citet{tolstoycetus00} 
is dominated by stars redder than the RR Lyrae instability strip. 
However, even if sparsely populated, a blue HB emerges clearly 
from our photometry. It is difficult to estimate how extended it is on the 
blue side, because it merges with the sequence of blue candidate MS stars/BSs.
We also note that there are a small number of objects on both the blue ($m_{F814W}
<$ 23 mag, $(m_{F475W} - m_{F814W}) \sim$ 1) and red side of the RGB. These are 
most likely foreground field stars.

\begin{figure*}[t]
\epsscale{1.0}
%\plotone{cetus_iso_allf.eps}
\plotone{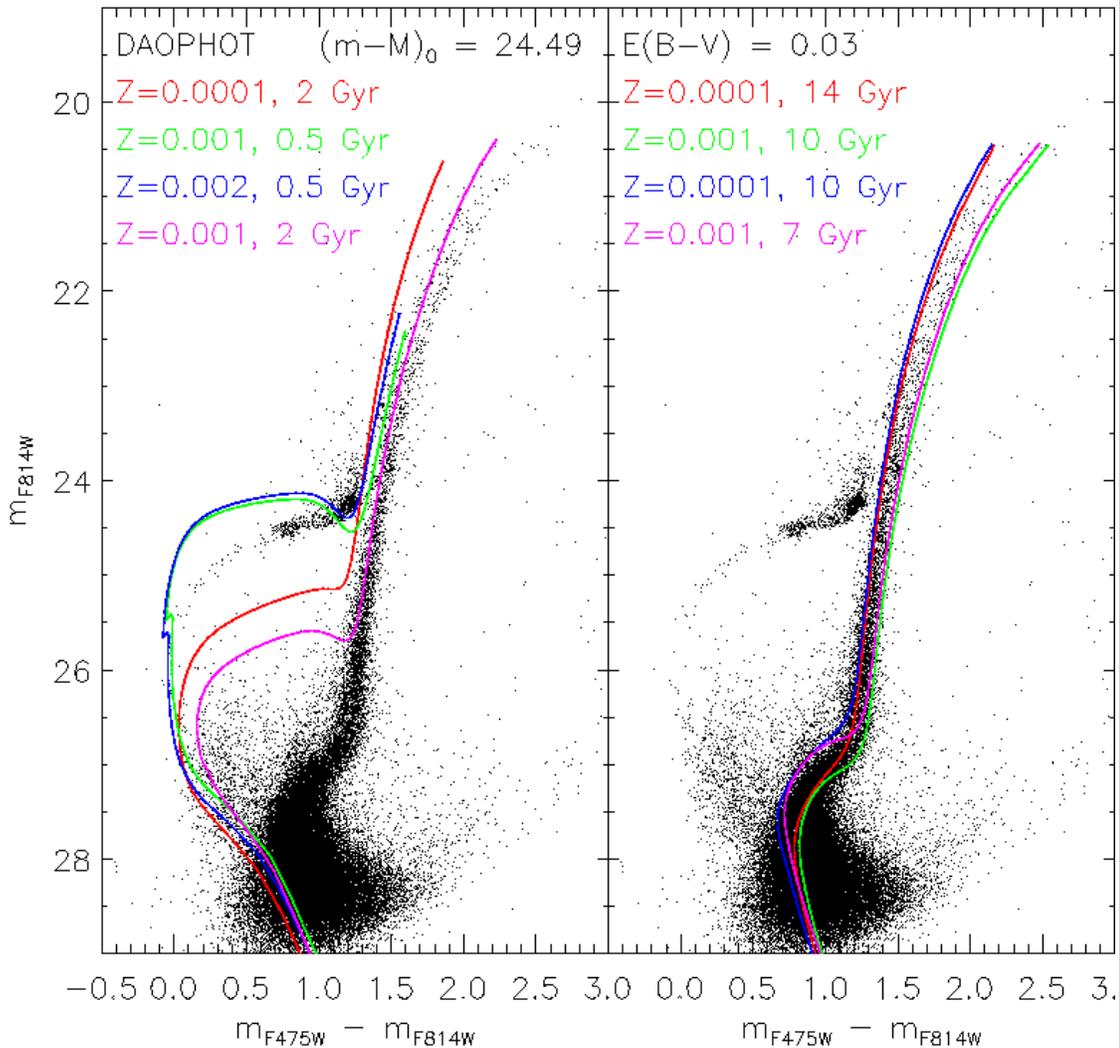}
\caption{Color-magnitude diagram, calibrated in the VEGAMAG photometric 
system, with overimposed some selected isochrones from the BaSTI database, 
shifted according to the labeled distance modulus and 
reddening values. Age and metallicity of the isochrones are also 
labeled in the figure. 
\label{fig:cmd_iso}}
\end{figure*}

We can gain some insight into the Cetus stellar populations
through a comparison with theoretical isochrones, which allows us to estimate
some rough limits to the ages and metallicities of the stellar populations.
Fig. \ref{fig:cmd_iso} presents the superposition of selected 
isochrones from the BaSTI database \citep{pietrinferni04}, for the indicated 
ages and metallicities. For simplicity, we adopted the DAOPHOT 
photometry, but the conclusions are independent of the photometry set.
The distance, $(m-M)_0=24.49$, was estimated using the mean magnitude of the 
RR Lyrae stars \citep{bernardcetustucana}\footnotemark[18].
An attempt to use the TRGB showed that the number of stars near 
the tip is too small to derive a precise estimate of the distance modulus.
However, the excellent agreement between the TRGB of the isochrones with the
brightest stars of the RGB in Fig. \ref{fig:cmd_iso} shows that this distance
is very reliable. The extinction $E(B-V) = 0.03$ was taken from 
\citet{schlegel98}, and was transformed to A$_{F475W}$ and A$_{F814W}$ following
\citet{bedin05}.

\footnotetext[18]{The value adopted in the SFH derivation, 24.49, is slightly 
different from the final value of 24.46 $\pm$ 0.12 
given in \citep{bernardcetustucana}.
However, such a small difference has negligible impact on the derived SFH.}

%%%%%%%%%%%%%%%%%%%%%%%%%%%%%%%%%%%%%%%%%%%%%%%%%%%%%%%%%%%%%%%%%%%%%%
%%%%%%%%%%%%%%%%%%%%%%%%%%%%% FIG 6 %%%%%%%%%%%%%%%%%%%%%%%%%%%%%%%%%%

%%%%%%%%%%%%%%%%%%%%%%%%%%%%%%%%%%%%%%%%%%%%%%%%%%%%%%%%%%%%%%%%%%%%%%
%%%%%%%%%%%%%%%%%%%%%%%%%%%%%%%%%%%%%%%%%%%%%%%%%%%%%%%%%%%%%%%%%%%%%%

The left panel of Fig. \ref{fig:cmd_iso} shows the comparison with young isochrones. 
The blue edge of the blue plume can be well matched with a metal-poor (Z = 0.0001),
relatively young (2 Gyr) isochrone (red line). A similarly good agreement with isochrones
of higher metallicity (Z = 0.001, 0.002, green and blue lines) is obtained only by 
reducing the age considerably (0.5 Gyr), but the lack of observed bright 
TO stars makes this interpretation unlikely. An increase in the age for the 
Z = 0.001 isochrone (pink line) cannot account for the observationally well 
measured blue portion of this feature.
This might suggest that this is a sequence of BSs rather than
a truly young MS (see also \S \ref{sec:bs}).

The right panel of the same figure shows the comparison with isochrones of age 
ranging from 7 to 14 Gyr, and metallicities 
Z = 0.0001 and Z = 0.001. It is interesting to note that both the TO and 
RGB phases are well confined in this metallicity range. In particular, the blue 
edge of the RGB seems to correspond to the lower metallicity, while the 
red edge is well constrained by a metallicity 10 times higher, with negligible 
dependence on age. The situation around the TO is more complicated. At fixed 
metallicity, the oldest isochrone matches the faint red edge of the TO
reasonably well, while the youngest nicely delimits the brighter and bluer one. 
Therefore, 
both the faint red and the bright blue edges can be properly represented using 
either an older, more metal-poor, or a slightly younger, more metal-rich isochrone.

However, from this simple analysis, we can not set strong constraints. 
A possible interpretation is that Cetus
consists of a dominant population of age 10 Gyr, with a small age range and with 
metallicity in the range 0.0001 $<$ Z $<$ 0.001. Alternatively, and equally consistent
with this simple analysis, is a scenario with both age and metallicity spreads.
Therefore, this simplistic analysis is only helpful to put loose constraints on the 
expected populations, in terms of ages and metallicities, but does 
not allow us to draw any quantitative conclusions on the age and metallicity
distribution of the observed stars.  For that, a more sophisticated analysis is 
required such as the one presented in \S \ref{sec:methods} below.

%%%%%%%%%%%%%%%%%%%%%%%%%%%%%%%%%%%%%%%%%%%%%%%%%%%%%%%%%%%%%%%%%%%%%%
%%%%%%%%%%%%%%%%%%%%%%%%%%%%% FIG 7 %%%%%%%%%%%%%%%%%%%%%%%%%%%%%%%%%%

%%%%%%%%%%%%%%%%%%%%%%%%%%%%%%%%%%%%%%%%%%%%%%%%%%%%%%%%%%%%%%%%%%%%%%
%%%%%%%%%%%%%%%%%%%%%%%%%%%%%%%%%%%%%%%%%%%%%%%%%%%%%%%%%%%%%%%%%%%%%%

	\subsection{The RGB bump}

\begin{figure}
\epsscale{1.2}
%\plotone{bumpolo.eps}
\plotone{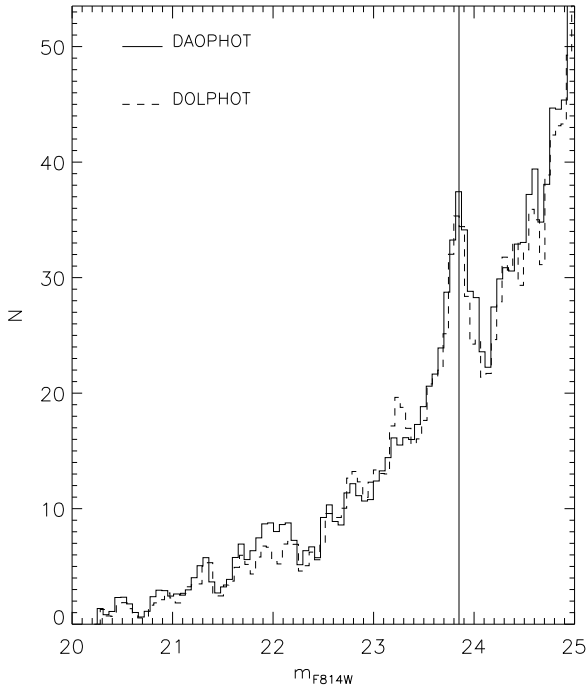}
\caption{Luminosity function of the Cetus RGB, for both the DAOPHOT 
(solid line) and DOLPHOT (dashed) photometry. The vertical line shows 
the position of the RGB bump.
\label{fig:bumpolo}}
\end{figure}

The RGB bump is a feature commonly observed in old stellar systems
\citep[e.g.,][]{riello03}, whose physical origin is relatively well understood
\citep{thomas67, iben68}. When a low-mass 
star ascends the RGB, the H-burning shell moves outward and reaches 
the chemical discontinuity left by the convective 
envelope at the time of its greatest depth. This forces a rearrangement of the stellar structure and a 
temporary decrease in luminosity, before the star resumes evolving toward 
higher luminosities and lower effective temperatures. From the observational 
point of view, since the star crosses the same luminosity interval three 
times, it produces an accumulation of objects at one point in the luminosity 
function of the RGB. Fig. \ref{fig:bumpolo} shows the observed Cetus RGB
luminosity function for both sets of photometry. The sample stars
were selected in a box closely enclosing the RGB, in order to minimize
the contribution of AGB and HB stars. The RGB bump clearly 
shows up, and fitting a Gaussian profile indicates a peak magnitude of 
$m_{F814} \sim$ 23.84 $\pm 0.08$, with excellent agreement between the 
two photometry sets.

%%%%%%%%%%%%%%%%%%%%%%%%%%%%%%%%%%%%%%%%%%%%%%%%%%%%%%%%%%%%%%%%%%%%%%
%%%%%%%%%%%%%%%%%%%%%%%%%%%%% FIG 8 %%%%%%%%%%%%%%%%%%%%%%%%%%%%%%%%%%

%%%%%%%%%%%%%%%%%%%%%%%%%%%%%%%%%%%%%%%%%%%%%%%%%%%%%%%%%%%%%%%%%%%%%%
%%%%%%%%%%%%%%%%%%%%%%%%%%%%%%%%%%%%%%%%%%%%%%%%%%%%%%%%%%%%%%%%%%%%%%

\section{Derivation of the star formation history of Cetus}\label{sec:methods}

In this section we present the various approaches used to derive
the SFH of Cetus. In particular, we adopted the two sets of photometry 
previously described, two stellar evolution libraries, 
BaSTI\footnotemark[19] \citep{pietrinferni04}
and Padova/Girardi\footnotemark[20] 
\citep{girardi00, marigo08}, and three different SFH codes: IAC-pop 
\citep{iacpop}, MATCH \citep{dolphin02}, and COLE \citep{skillman03}. 
We did not explore all the possible combinations of photometry/library/method, 
but we derived different solutions useful for comparison purposes, and to 
search for possible systematics. In particular, IAC-pop was applied to both sets 
of photometry and libraries, while MATCH and COLE were tested with the DOLPHOT 
photometry in combination with the Girardi library.

\footnotetext[19]{\itshape http://www.oa-teramo.inaf.it/BASTI}
\footnotetext[20]{\itshape http://pleiadi.oapd.inaf.it/}

	\subsection{The IAC method}\label{sec:method_iacpop}

This method has been developed in recent years by members of the {\itshape 
Instituto de Astrof\'{i}sica de Canarias} (A.A., S.H., C.G.), and
now includes different independent modules. Each software package has been presented
in different papers (IAC-star: \citealt{iacstar}, MinnIAC: Hidalgo et al., 
in prep., %\citet{hidalgolgs3},
IAC-pop: \citealt{iacpop}), which are the reference publications for the 
interested reader and include appropriate references\footnotemark[21]. 
In the following we give only a practical overview of 
the parameters adopted to derive the SFH of Cetus.

\footnotetext[21]{IAC-star and IAC-pop
are freely available at http://iac-star.iac.es and 
http://www.iac.es/galeria/aaj/iac-pop\_eng.htm, respectively.}

The entire procedure can be divided into five steps.
Note that in the following we will use ``synthetic CMD"  to designate
the CMD generated with IAC-star, and we will call it a ``model CMD" if the
observational errors have been simulated in the synthetic CMD.

{\itshape 1) The synthetic CMD:} it is created using IAC-star
\citep{iacstar}. We calculated a synthetic CMD with 8,000,000
stars from the inputs described in the following. We have verified 
that, with this large number of stars, no systematic effects
in the derived SFH of mock galaxies---such as those discussed in
\citet{noel09}---were observed, and so no corrections such as
those described in that paper are necessary.
The requested inputs are:

\begin{itemize}
\item a set of theoretical stellar evolution models. We adopted the 
BaSTI \citep{pietrinferni04} and Padova/Girardi \citep{girardi00} 
libraries;
\item a set of bolometric corrections to transform the theoretical stellar 
evolution tracks into the ACS camera photometric system. We applied the 
same set, taken from \citet{bedin05}, to both libraries.
\item the SFR as a function of time, $\psi(t)$. We 
used a constant SFR  in the age range $0 < t < 15 $ Gyr;
\item No {\itshape a priori} age-metallicity relation Z(t) is adopted:
the stars of any age have metallicities uniformly 
distributed in the range 0.0001$<$ Z $<$0.005. We stress that the 
age and metallicity intervals are deliberately selected to be 
{\itshape wider} than the ranges expected in the solution. This 
is done to ensure that no information is lost, and to provide the code with  
enough degrees of freedom to find the best possible solution;
\item the initial mass function (IMF), taken from \citet{kroupa02}. 
This is expressed with the formula $N(m)~dm = m^{-\alpha}~dm$, where $\alpha$
 = 1.3 for stars with mass smaller than 0.5 M$_\odot$, and $\alpha$ = 2.3
for stars of higher mass. Different values of both exponents 
have been tested with IC~1613 and LGS~3. The results,
presented in Skillman et al. (in prep.), show that the best 
solutions are obtained with values compatible with the Kroupa IMF;
\item the binary fraction, $\beta$, and the relative mass 
distribution of binary stars, $q$
\footnotemark[22]. We tested six values of 
the binary fraction, from 0\% to 100\%, in steps of 20\%, with fixed
$q>$ 0.5; in the following analysis we use 40\% and discuss the impact
of different assumptions in the Appendix I;
\item the assumed mass loss during the RGB phase follows the
empirical relation by \citet{reimers75}, with efficiency $\eta = 0.35$.
\end{itemize}

\footnotetext[22]{This limit is set for practical purposes, since binaries 
with higher mass ratios do not affect the distribution of stars in the CMD 
\citep[see][]{hurley98}. A 40\% fraction of binaries with this approach 
implies a larger total binary fraction. See \citet{gallart99} for a 
discussion of this and related issues.}

\begin{figure}
\epsscale{1.5}
%\plottwo{h1.eps}{h2.eps}
\plottwo{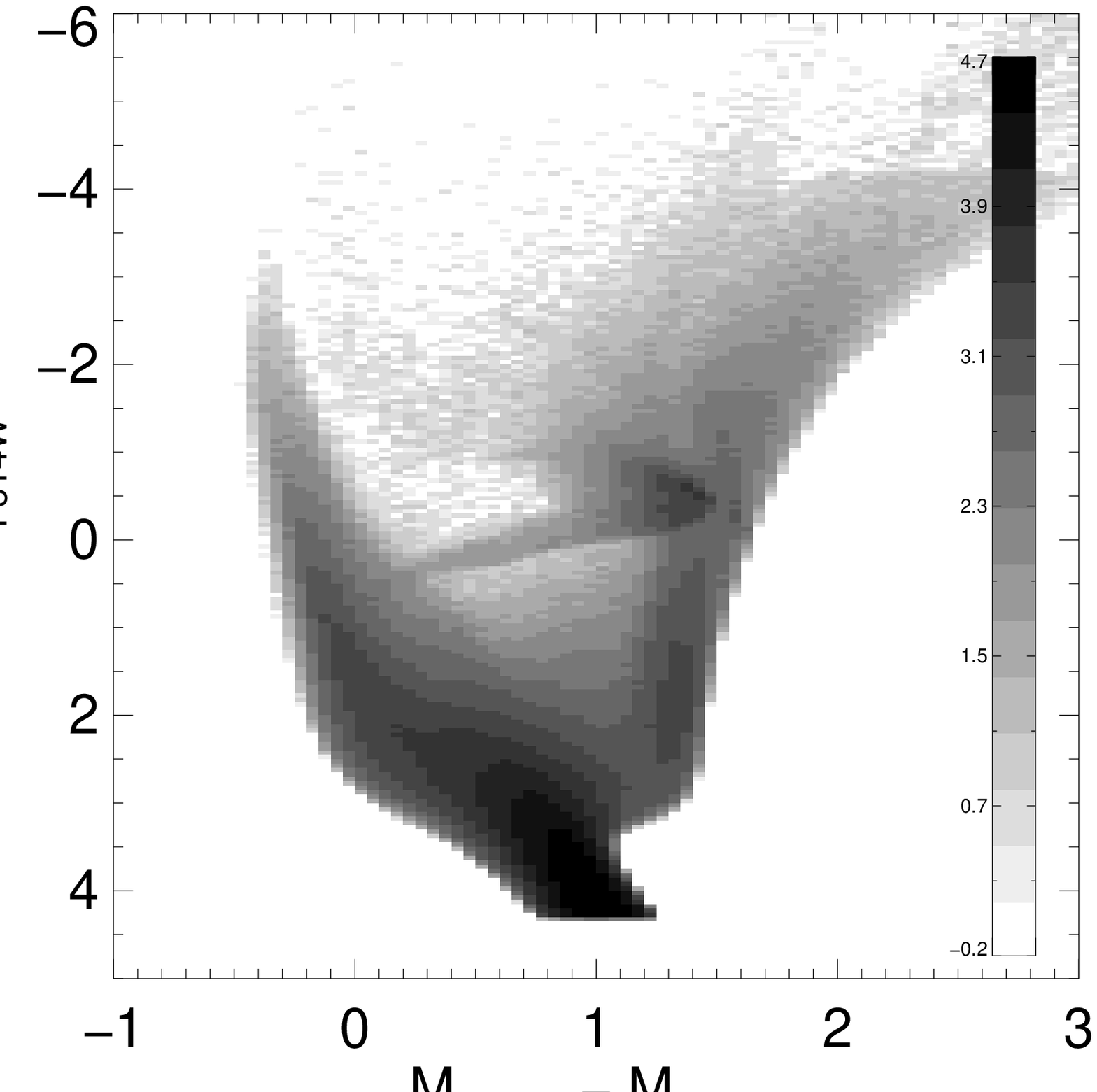}{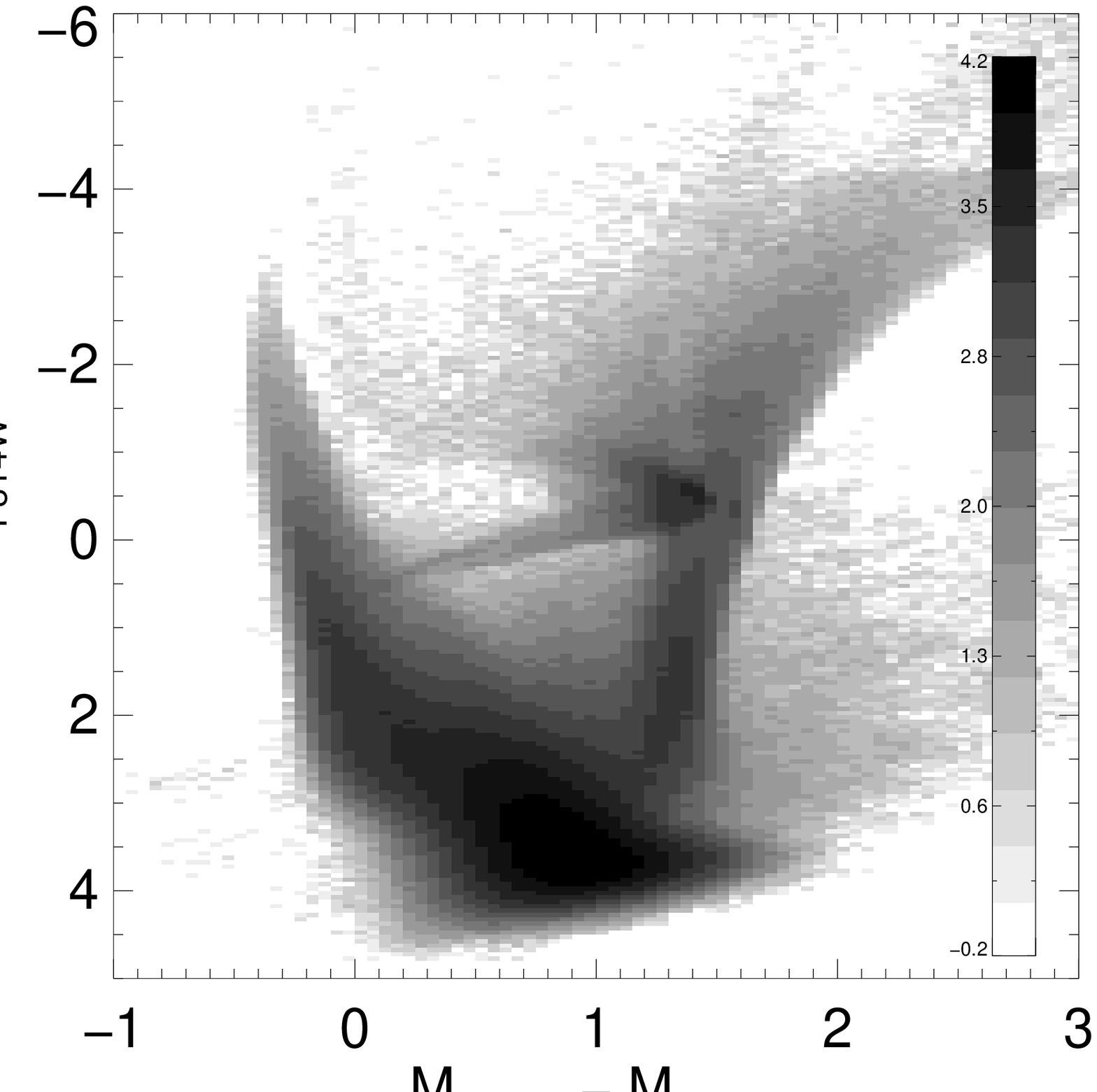}
\caption{Hess diagram of the 8 $\times 10^6$ stars synthetic CMD generated 
with IAC-star (upper) and after simulating the observational errors (lower).
\label{fig:simu}}
\end{figure}

{\itshape 2) The error simulation:} The code to simulate the observational 
errors in the synthetic CMD, called {\itshape obsersin}, has been developed
following \citet{gallart96}, and is described in Hidalgo et al., in prep. %\citet{hidalgolgs3}.
It takes into account the incompleteness and photometric errors due to 
crowding using an empirical approach, taking the
information from the completeness test described in \S \ref{sec:crow_dao}
and \S \ref{sec:crow_dol}.
 This is a fundamental step because 
the distribution of stars in the observed CMD is strongly modified
from the actual distribution due to the observational errors,
particularly at the faint magnitude level near the old TO, which is 
where most of the information on the most ancient star formation 
is encoded.
Fig. \ref{fig:simu} shows an example of synthetic CMD as 
generated by IAC-star {\itshape (upper panel)}, and after the 
dispersion according to the observational errors of Cetus 
{\itshape (lower panel)}.

%%%%%%%%%%%%%%%%%%%%%%%%%%%%%%%%%%%%%%%%%%%%%%%%%%%%%%%%%%%%%%%%%%%%%%
%%%%%%%%%%%%%%%%%%%%%%%%%%%%% FIG 9 %%%%%%%%%%%%%%%%%%%%%%%%%%%%%%%%%%

%%%%%%%%%%%%%%%%%%%%%%%%%%%%%%%%%%%%%%%%%%%%%%%%%%%%%%%%%%%%%%%%%%%%%%
%%%%%%%%%%%%%%%%%%%%%%%%%%%%%%%%%%%%%%%%%%%%%%%%%%%%%%%%%%%%%%%%%%%%%%

{\itshape 3) Parameterization:} MinnIAC 
%(Hidalgo et al., in prep.)
%\citep{hidalgolgs3} 
is a suite 
of routines developed specifically to accomplish two main purposes:
first, an efficient sampling of the parameter space to obtain a set of solutions
with different parametrizations of the CMD, and second, estimation of
the average SFH from the different solutions that best
represent the observations.
In our case, ``parameterization" of the CMD means two things:
{\itshape i)} defining the age and metallicity bins that fix the simple 
populations and {\itshape ii)} defining a grid of boxes covering 
critical evolutionary phases in the CMDs, where the stars of the observed 
and model CMD are counted.

%%%%%%%%%%%%%%%%%%%%%%%%%%%%%%%%%%%%%%%%%%%%%%%%%%%%%%%%%%%%%%%%%%%%%%
%%%%%%%%%%%%%%%%%%%%%%%%%%%%% FIG 10 %%%%%%%%%%%%%%%%%%%%%%%%%%%%%%%%%

\begin{figure}
\epsscale{1.20}
%\plotone{bundles.eps}
\plotone{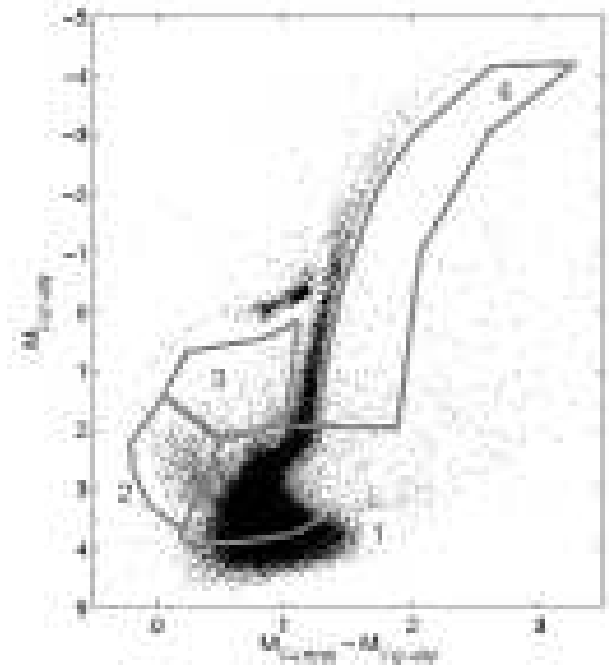}
\caption{Cetus observed CMD, transformed to the absolute plane assuming 
(m-M)$_0$ = 24.49 and E(B$-$V)=0.03, with the location of the four bundles
superimposed.
\label{fig:bundles}}
\end{figure}

%%%%%%%%%%%%%%%%%%%%%%%%%%%%%%%%%%%%%%%%%%%%%%%%%%%%%%%%%%%%%%%%%%%%%%
%%%%%%%%%%%%%%%%%%%%%%%%%%%%%%%%%%%%%%%%%%%%%%%%%%%%%%%%%%%%%%%%%%%%%%
%
The basic age and metallicity bins defining the simple populations used
in this work are: \\
{\itshape age :} [1.5 2 3 ... 13 14 (15)]*10$^9$ years \\
{\itshape metallicity :} [0.1 0.3 0.5 0.7 1.0 1.5 2.0]*10$^{-3}$ \\

Note that model stars younger than 1.5~Gyrwere not included because there 
is no evidence of their presence in Cetus.  These numbers define 
the boundaries of the bins, not their centers. Therefore, they fix 
$15 \times 6 = 90$ simple populations.
The choice of a 1 Gyr bin size at all ages was adopted after testing a range
of bin widths. It was found that smaller bins do not increase the age resolution;
rather they increase the noise in the solution. The selected size of 1 Gyr
is the optimal compromise to fully exploit the data, within the limits imposed
by the observational errors and the intrinsic age resolution in the CMD
at old ages.

The parameterization of the CMD relies on the concept of {\itshape bundles} 
(see \citealt{iacpop} and Fig. \ref{fig:bundles}), that is, macro-regions on 
the CMD that can be sub-divided into boxes using different appropriate 
samplings. The number of boxes in the bundle determines the weight that 
the region has for the derived SFH. This is an efficient and flexible 
approach for two reasons. First, it allows a 
finer sampling in those regions of the CMD where the models are less 
affected by the uncertainties in the input physics (MS, SGB). Second, 
the dimensions of the boxes can be increased, and the impact on the
solution decreased, in those regions where the small number of observed 
stars could introduce noise due to small number statistics or where we
are less confident in the stellar evolution predictions. We optimized 
the values for this work using four bundles, with the 
following box sizes. The smallest boxes ($0.01~mag \times 0.2~mag$, 
N1 $\sim$ 900 boxes) sample the lower (old) MS, and the SGB 
(bundle 1). The region of the candidate BSs (bundle 2) is mapped with
slightly bigger boxes ($0.03~mag \times 0.5~mag$, N2 $\sim$ 90 boxes),
to minimize the number of boxes with zero or few stars.
Bundle 3 samples the region between the BS sequence and the RGB.
No bundles include the stars in evolved evolutionary phases,
and in particular we excluded both the RGB and the HB from our analysis.
The reasons are many. First, the physics
governing these evolutionary phases is more uncertain than that
describing the MS, and differences between stellar libraries
are more important in these evolved phases \citep{gallartreview}.
Moreover, the details of the HB morphology also depend on highly
unknown factors like the mass loss during the RGB phase. However,
despite not including the RGB stars, we routinely adopt a bundle
which is {\itshape redder} than the RGB (bundle 4 in Fig. \ref{fig:bundles}).
In order to aid in setting a mild constraint on the upper limit
of the metallicity of the RGB Cetus stars, this was drawn at
the red edge of the RGB. It includes metal rich stars present
in the model, but very few objects of the observed CMD.
Both bundles 3 and 4 contain boxes of $1.5~mag \times 1.0~mag$
(N3 = 7, N4 = 4). Extensive tests have shown that including a bundle
on the RGB did not improve the solution, nor produce a solution significantly
different. Rather, the $\chi^2$ of the solution increases substantially
and likely spurious populations appear in the 3D histogram, such as
old and metal-rich stars.

For a given set of input parameters, the simple populations plus 
the grid of bundles+boxes, MinnIAC counts the stars in the boxes 
for both the observed and all the simple populations in the model 
CMD. A simple table with these star counts is the input information
to run IAC-pop. However, our numerous tests disclosed that this 
approach can be significantly improved by adopting a series of 
slightly different sets of input parameters, used to derive many
solutions and then calculating the mean SFH. This ``dithering approach",
which is the distinctive feature of the code, significantly reduces 
the fluctuations associated with the adopted sampling (both in terms 
of simple populations and boxes) of each individual solution 
(Hidalgo et al., in prep.). %\citep{hidalgolgs3}.

In the case of the Cetus SFH, the age and metallicity bins are 
shifted three times, each time by an amount equal to 30\% of the 
bin size, with four different configurations: {\em i)} moving the 
age bin toward increasing age (with fixed metallicity); 
{\em ii)} moving the metallicity bin toward increasing metallicity 
(with fixed age); {\em iii)} moving both bins toward increasing 
values; and {\em iv)} moving toward decreasing age and increasing 
metallicity. These 12 different sets of simple populations are 
used twice, shifting the boxes a fraction of their size across the CMD.

Moreover, to take into account the uncertainties associated with the
distance and reddening estimates, together with all the hidden 
systematics possibly affecting the zero points of the photometry, 
MinnIAC also repeats the whole procedure after shifting the observed 
CMD (not the model) a number of steps in both color and magnitude. The 
bundles are correspondingly shifted. 
For this project, we adopted an initial grid of 25 positions, 
shifting the observed CMD in magnitude by [$-$0.15, $-$0.075, 0, 
+0.075, +0.15] mag, and in color by [$-$0.06, $-$0.03, 0, +0.03, 
+0.06] mag. Since we calculated 24 solutions for each node of 
this grid, we derived a total of 600 solutions.

We stress that the position of the best $\chi^2_\nu$ in the magnitude 
and color grid is {\itshape not} intended for estimates of distance or 
reddening, since photometry zero points or model systematics also 
play a role. However, we think it is reasonable to take the best 
agreement between model and observations in that particular point 
of the ($\delta_{col}$, $\delta_{mag}$) grid as pointer of the best
``absolute" SFH solution.

{\itshape 4) Solving:} IAC-pop \citep{iacpop} is a code designed
to solve the SFH of a resolved stellar system. It uses only the information
of the star counts in the different boxes across the CMD.
Therefore, it calculates which combination of partial models gives
the best representation of the stars in the observed CMD, using the
modified $\chi^2_\nu$ merit-function introduced by \citet{mighell99}.
Note that IAC-pop solves the SFH using age and metallicity
as independent variables, without any assumption on the age-metallicity
relation. IAC-pop was run independently on each different 
initial configuration created with MinnIAC. 

{\itshape 5) Averaging:} A posteriori, MinnIAC was also used to read
the individual solutions, and calculate the mean best SFH.

	\subsection{The MATCH method}\label{sec:method_match}

The basis of the MATCH method of measuring SFHs \citep{dolphin02} is
that it aims at minimizing the difference between the observed CMD 
and synthetically generated CMDs. The SFH that produces the best-fit 
synthetic CMD is the most likely SFH of the observed CMD. To 
determine the best-fit CMD, the MATCH method uses a Poisson maximum 
likelihood statistic to compare the observed and synthetic CMDs.
For specific details on how this is implemented in MATCH, 
see \citet{dolphin00}.

To appropriately compare the synthetic and observed CMDs, we binned them 
into Hess diagrams by 0.10 mags in m$_{F475W}$ and 0.05 mags in m$_{F475W}$
$-$ m$_{F814W}$. For consistency, we chose parameters to create synthetic 
CMDs that closely matched those used in the IAC-pop code presented 
in this paper: \\
$\bullet$ a power law IMF with $\alpha$ $=$ 2.3 from 0.1 to 100 M$_\odot$; \\
$\bullet$ a binary fraction $\beta$ $=$ 0.40; \\
$\bullet$ a distance modulus of 24.49, A$_{F475W}$ $=$ 0.11; \\
$\bullet$ the stellar evolution libraries of \citet{marigo08},
which are the basic Girardi models with updated AGB evolutionary 
sequences.

The synthetic CMDs were populated with stars over a time range of
$\log{t}$ $=$ 6.6 to $\log{t}$ $=$ 10.20, with a uniform bin size of 0.1.
Further, the program was allowed to solve for the best fit metallicity per 
time bin, drawing from a range of [M/H] = $-$2.3 to 0.1, where M canonically
represents all metals in general. The depth of the photometry used for 
the SFH was equal to 50\% completeness in both m$_{F475W}$ and m$_{F814W}$.

To quantify the accuracy of the resultant SFH, we tested for both 
statistical errors and modeling uncertainties. To simulate possible 
zero point discrepancies between isochrones and data, we constructed 
SFHs adopting small offsets in distance and extinction from the best
fit values. The rms scatter between those solutions is a reasonable proxy
for uncertainty in the stellar evolution models and/or photometric 
zero-points. Statistical errors were accounted for by solving 
fifty random realizations of the best fit SFH. Final error bars are calculated
summing in quadrature the statistical and systematic errors.

%%%%%%%%%%%%%%%%%%%%%%%%%%%%%%%%%%%%%%%%%%%%%%%%%%%%%%%%%%%%%%%%%%%%%%
%%%%%%%%%%%%%%%%%%%%%%%%%%%%% FIG 11 %%%%%%%%%%%%%%%%%%%%%%%%%%%%%%%%%

\begin{figure*}
\epsscale{1.0}
%\resizebox{15truecm}{15truecm}{\includegraphics[clip=true]{external.eps}}
\resizebox{15truecm}{15truecm}{\includegraphics[clip=true]{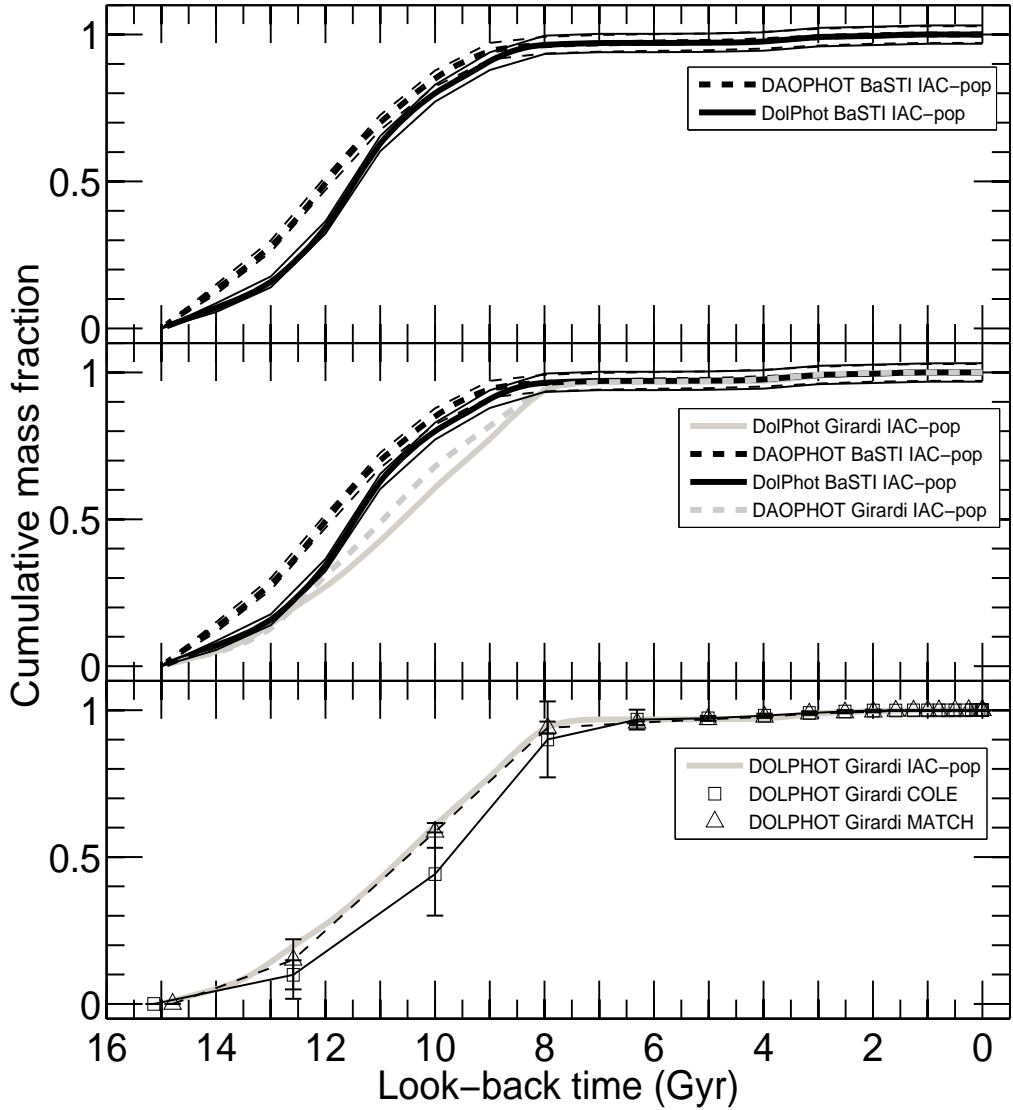}}
\caption{Summary of the different results applying different 
techniques to derive the SFH. The cumulative mass fraction
as a function of time is shown. {\itshape Top:} Comparison of the SFH
derived with the MinnIAC/IAC-pop method, using the BaSTI library and
the DAOPHOT and DOLPHOT photometry. The DAOPHOT SFH is 
overall less than 1 Gyr older then the DOLPHOT one. {\itshape Middle: } 
The same solutions are compared to analogous ones derived using the 
Padova/Girardi library. These solutions are systematically younger 
than the BaSTI solutions, particularly at old epochs for the DAOPHOT
photometry. {\itshape Bottom:} Comparison of two different SFH codes,
applied to the DOLPHOT photometry in combination with the Padova/Girardi
library.
\label{fig:external}}
\end{figure*}

%%%%%%%%%%%%%%%%%%%%%%%%%%%%%%%%%%%%%%%%%%%%%%%%%%%%%%%%%%%%%%%%%%%%%%
%%%%%%%%%%%%%%%%%%%%%%%%%%%%%%%%%%%%%%%%%%%%%%%%%%%%%%%%%%%%%%%%%%%%%%

	\subsection{The COLE method}\label{sec:method_cole}

The third method we applied to measure the SFH of Cetus is a simulated
annealing algorithm applied to our DOLPHOT photometry and the Padova/Girardi
library. The code is the same as that applied to the galaxies Leo~A 
\citep{cole07} and IC~1613 \citep{skillman03}, and details of the 
method can be found in those references. The technique treats the 
task as a classic inverse problem by finding the SFH that
maximizes the likelihood, based on Poisson statistics, that the 
observed distribution of stars in the binned CMD (Hess diagram) could 
have been drawn from the model.  

The code uses computed stellar evolution and atmosphere models that give 
the stellar colors and magnitudes as a function of time for stars of a 
wide range of initial masses and metallicities. The isochrone tables \citep{marigo08} 
for each (age, metallicity) pair are shifted by the appropriate distance 
and reddening values, and transformed into a discrete color-magnitude-density 
distribution $\varrho$ by convolution with an initial mass function 
\citep[see ][, their Tab. 1]{chabrier03}.
and the results of the artificial star tests simulating observational uncertainties
and incompleteness.  The effects of binary stars on the observations are 
simulated by adopting the binary star frequency and mass ratio distribution 
from the solar neighborhood, namely
35\% of stars are single, 46\% are parameterized as ``wide" binaries, i.e., 
the secondary mass is uncorrelated with the primary mass, and is drawn from
the same IMF as the primary, and finally 19\% of stars are parameterized as 
``close" binaries, i.e., the probability distribution function of the mass 
ratio is flat.% \citep[for details see][]{cole07}.
The progeny of dynamical interactions between binaries (e.g., 
blue stragglers) are not modelled.  The history of the 
galaxy is then divided into discrete age bins with approximately 
constant logarithmic spacing in order to take advantage of the 
increasing time resolution allowed by the data at the bright 
(young) end.  The star formation history is then modelled as 
the linear combination of $\psi$(t,Z) that best matches the data 
by summing the values of $\varrho$ at each point in the Hess diagram.  
The best fit is found via a simulated annealing technique that 
converges slowly on the maximum-likelihood solution without becoming 
trapped in local minima (such as those produced by age-metallicity 
degeneracy).

In our solution we considered only the DOLPHOT photometry. The critical 
distance and reddening parameters were held fixed at (m$-$M)$_0$ = 24.49 
and and E(B$-$V) = 0.03. We used 10 time bins ranging from 6.60 $\leq$ 
log(age/Gyr) $\leq$ 10.18. Isochrones evenly spaced by 0.2 dex
from $-2.3 \leq [M/H] \leq -0.9$ were used, considering only scaled-solar
abundances but with no further constraints on the age-metallicity relation.
The CMD was discretized into bins of dimension
0.10$\times$0.20 mag, and it is the density distribution in 
this Hess diagram that was fit to the models.  The error bars are determined
by perturbing each component of the best-fit SFH in turn and re-solving for
a new best fit with the perturbed component held fixed.  The size of the 
perturbation is increased until the new best fit is no longer within 
1$\sigma$ of the global best fit.

	\subsection{Comparison of the different solutions}\label{sec:sfh_comparison}

Fig. \ref{fig:external} summarizes the comparison of the solutions 
calculated using different sets of photometry, stellar evolution libraries, 
and SFH codes. 

The upper panel shows the cumulative mass fraction, as a function 
of time, based on the BaSTI library + IAC method applied to 
the two sets of photometry. The solution based on the DOLPHOT photometry 
is marginally younger than the DAOPHOT one. It shows slightly lower
star formation at older epochs, and a steeper increase around 12 Gyr ago. 
In either case Cetus completely stopped forming stars $\sim$ 8 Gyr ago.

The middle panel compares the same two curves with analogous ones 
obtained using the Girardi library. The effect of changing the library 
introduces a small systematic effect; both solutions with the Girardi 
library shift more star formation to slightly ($\le$ 1 Gyr) younger ages
compared to the solutions using the BaSTI library. This effect 
seems more evident at the oldest epochs in the case of the 
DAOPHOT solutions. 

The bottom panel shows the results of the three different SFH
reconstruction methods, applied to the same photometric catalog 
and with the same stellar evolution library. Since for the MATCH 
and COLE method we present the result of the best individual solution, and
not the average of many as in the case of the IAC method, the large dots
highlight the centers of the age bins adopted. Note the good 
agreement between the MATCH and COLE with the IAC solution, in spite of 
the higher time resolution of the latter.

For simplicity, we will now focus our analysis on the SFHs obtained with 
the IAC method and the BaSTI stellar evolution library. At this time, 
we favor using the BaSTI stellar evolution library because the 
input physics has been more recently updated, in comparison to 
the Girardi stellar evolution library \citep{pietrinferni04}.
Similarly, the IAC method will be used preferentially because
it was tailored to the needs of this project and affords
a greater degree of flexibility for the analysis (Hidalgo et al., in prep.).  %\citep{hidalgolgs3}.

\section{The SFH of Cetus} \label{sec:results}

In this section we present the details of the Cetus SFH. First, in 
\S \ref{sec:sfh_iacpop}, we describe the analysis we did to explore 
the possible systematics affecting the solution. We then describe our 
final solution in \S \ref{sec:sfh}.

	\subsection{Approaching the best solution}\label{sec:sfh_iacpop}

		\subsubsection{Exploring the zero points systematics}\label{sec:sfh_iacpop_a}

\begin{figure}
\epsscale{1.0}
%\resizebox{8truecm}{7truecm}{\includegraphics[clip=true]{chi2map.eps}}
\resizebox{8truecm}{7truecm}{\includegraphics[clip=true]{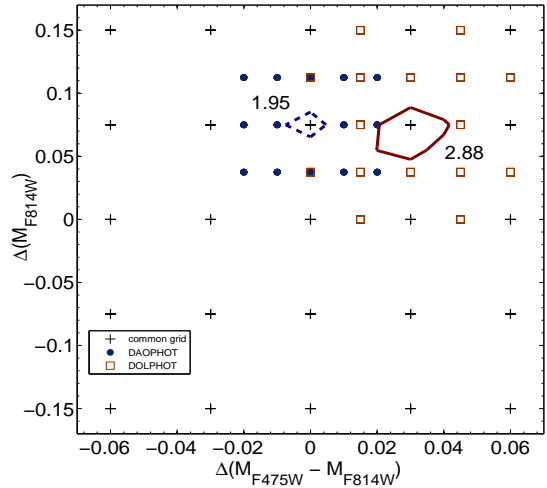}}
\caption{Summary of the solutions derived as a function of the shifts in
color and magnitude, $\delta_{col}$ and $\delta_{mag}$, applied to the
observed CMD. In each position of the grid we calculated 24 
solutions. Crosses mark the initial grid, common to both photometry sets. 
Filled circles and open squares show the sub-grid around the initial
minimum, for the DAOPHOT and DOLPHOT photometry set, respectively.
The two curves show the 1-$\sigma$ confidence area, defined
using the dispersion of the 24 individual solution around the mean $\chi^2_\nu$.
\label{fig:mapchi2}}
\end{figure}

Fig. \ref{fig:mapchi2} shows the grid of color and magnitude shifts
adopted to derive the IAC-pop solution. 
The {\itshape plus} signs mark the positions of the 25 initial points.
In each of these 25 positions we calculated the mean $\chi^2_\nu$,
averaging the $\chi^2_\nu$ of the 24 individual solutions. This allows us 
to study how the $\chi^2_\nu$ varies as a function of the magnitude
and color shifts. In particular, the minimum averaged $\chi^2_{\nu, min}$ 
identifies the position corresponding to the best solution,
which is calculated by MinnIAC averaging the corresponding 24 
individual solutions.

%%%%%%%%%%%%%%%%%%%%%%%%%%%%%%%%%%%%%%%%%%%%%%%%%%%%%%%%%%%%%%%%%%%%%%
%%%%%%%%%%%%%%%%%%%%%%%%%%%%% FIG 12 %%%%%%%%%%%%%%%%%%%%%%%%%%%%%%%%%

%%%%%%%%%%%%%%%%%%%%%%%%%%%%%%%%%%%%%%%%%%%%%%%%%%%%%%%%%%%%%%%%%%%%%%
%%%%%%%%%%%%%%%%%%%%%%%%%%%%%%%%%%%%%%%%%%%%%%%%%%%%%%%%%%%%%%%%%%%%%%

Because the initial color and magnitude shifts are relatively large,
we calculated more solutions on a finer grid with 14 positions 
around the identified minimum $\chi^2_\nu$ (Fig. \ref{fig:mapchi2}, full 
circles and open squares for the DAOPHOT and DOLPHOT solution, respectively).
Therefore, we obtained a total of (25+14)*24=936
solutions. This computationally expensive task was made possible by
the Condor workload management system \citep{condor}\footnotemark[23] 
available at the Instituto de Astrof\'isica de Canarias.
In the case of the DAOPHOT photometry, the minimum was confirmed 
at the position ($\delta_{col}$ ; $\delta_{mag}$) = (0.0 ; 0.075),
$\chi^2_{\nu,min}$ = 1.95, while in the case of the DOLPHOT photometry 
the minimum is located at ($\delta_{col}$ ; $\delta_{mag}$) = (+0.03, 0.075), 
$\chi^2_{\nu,min}$ = 2.88. The two curves around the minimum position 
of Fig. \ref{fig:mapchi2} represent the 1-$\sigma$ confidence area around 
the minimum, calculated as the dispersion around the mean of the 24 
$\chi^2$ of the individual solutions \citep{iacpop}. Note that the color 
difference between the $\chi^2_{\nu, min}$ of the two photometry sets is 
fully consistent with, and apparently compensates for, the color shift 
between the two photometry sets ($\sim 0.04$ mag). This indicates that 
the flexibility of our method is able to deal with subtle photometric 
calibration systematics, and possible stellar evolution model, distance, 
and reddening uncertainties.
\footnotetext[23]{http://www.cs.wisc.edu/condor/}

\begin{figure}
\epsscale{1.1}
%\plotone{cetus-sfhagemean40-dolphot-daophot.eps}
\plotone{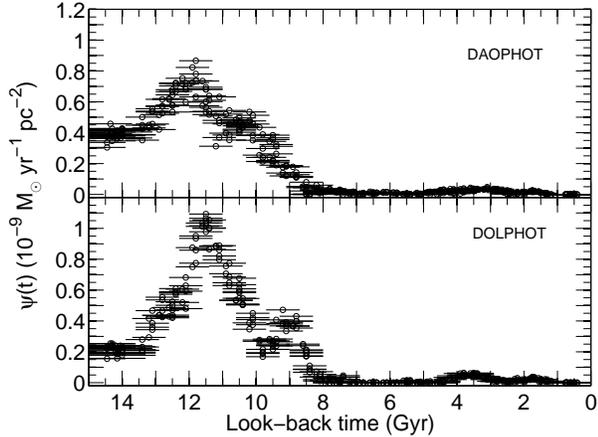}
\caption{Comparison of the 24 individual DAOPHOT and DOLPHOT solutions
at the best $\chi^2_\nu$ 
position, from both sets of photometry. Note the stability of the solution 
against small changes of the input bins adopted. Moreover, the plot puts 
into evidence the similarity of the solutions obtained from the DAOPHOT 
and DOLPHOT photometry. Note that the horizontal lines are not the error
bars, but represent the size of the age bin.
\label{fig:allbest_daodol}}
\end{figure}

Fig. \ref{fig:allbest_daodol} shows a direct comparison between the 24 
individual DAOPHOT and DOLPHOT solutions at the $\chi^2_{\nu, min}$ position. 
The horizontal lines show the size of the age bins, and do not
represent error bars. It is noteworthy that the spread of the 
24 solutions, derived using different input samplings, is quite small,  
suggesting the robustness of the mean SFH. Note also the good agreement 
between the solutions from the two photometry sets. The general trend 
is the same, with the main peak of star formation occurring within $\sim$ 
0.5 Gyr and, most importantly, characterized by the same duration. 

%%%%%%%%%%%%%%%%%%%%%%%%%%%%%%%%%%%%%%%%%%%%%%%%%%%%%%%%%%%%%%%%%%%%%%
%%%%%%%%%%%%%%%%%%%%%%%%%%%%% FIG 13 %%%%%%%%%%%%%%%%%%%%%%%%%%%%%%%%%

%%%%%%%%%%%%%%%%%%%%%%%%%%%%%%%%%%%%%%%%%%%%%%%%%%%%%%%%%%%%%%%%%%%%%%
%%%%%%%%%%%%%%%%%%%%%%%%%%%%%%%%%%%%%%%%%%%%%%%%%%%%%%%%%%%%%%%%%%%%%%

%
%%%%%%%%%%%%%%%%%%%%%%%%%%%%%%%%%%%%%%%%%%%%%%%%%%%%%%%%%%%%%%%%%%%%%%
%%%%%%%%%%%%%%%%%%%%%%%%%%%%% FIG 14 %%%%%%%%%%%%%%%%%%%%%%%%%%%%%%%%%
%
\begin{figure}
\epsscale{1.1}
%\plotone{cfr135.eps}
\plotone{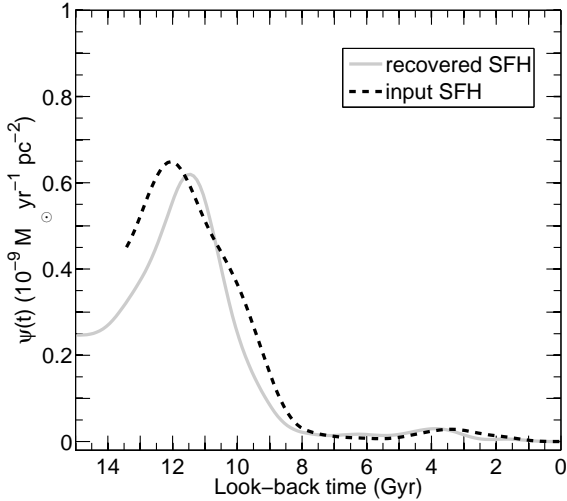}
\caption{Results  of a test designed to investigate the significance of 
the oldest $\psi(t)$ in our solutions. Dashed line: input SFH,  adopted 
as our Cetus solution, but removing all stars older than 13.5 Gyr. Solid 
line: recovered SFH when using a model CMD that contains stars as old as 
15 Gyr. A low level of star formation is recovered for ages older than 
13.5 Gyr, when there should be none.
\label{fig:135}}
\end{figure}
%
%%%%%%%%%%%%%%%%%%%%%%%%%%%%%%%%%%%%%%%%%%%%%%%%%%%%%%%%%%%%%%%%%%%%%%
%%%%%%%%%%%%%%%%%%%%%%%%%%%%%%%%%%%%%%%%%%%%%%%%%%%%%%%%%%%%%%%%%%%%%%
%

		\subsubsection{The age of the oldest stars}\label{sec:sfh_iacpop_b}

Another important aspect to be addressed is the occurrence of star
formation at epochs older than 14 Gyr. As discussed in 
\S \ref{sec:method_iacpop}, we did not impose 
any strong constraints on the ages of the oldest stars in the model used 
to derive the SFH of Cetus. That is, we let the code search for the best 
solution using model stars as old as 15 Gyr, significantly older than the 
age of the Universe commonly accepted after the WMAP experiment \citep{bennett03}. 
Interestingly enough, our solution shows very low star formation rate for 
epochs older than 14 Gyr, with a steady increase toward a peak at $\sim$ 
12 Gyr. \footnotemark[25]

\footnotetext[24]{It is also worth noting that none of the stellar evolution libraries  
implemented in IAC-star accounts for the occurrence of atomic diffusion.
Atomic diffusion has the effect of reducing the evolutionary lifetimes 
during the central H-burning stage of low-mass stars, resulting
in a reduction of the stellar age - at the  
oldest ages - by about 0.7 Gyr \citep{castellani97, cassisi99}.
Therefore, in the present analysis, an age estimate of $\sim14 
$~Gyr would correspond to an age of about 13.3 Gyr if models accounting 
for atomic diffusion were to be adopted. Both are in very good  
agreement with current determination of the age of the Universe.}

%
%%%%%%%%%%%%%%%%%%%%%%%%%%%%%%%%%%%%%%%%%%%%%%%%%%%%%%%%%%%%%%%%%%%%%%
%%%%%%%%%%%%%%%%%%%%%%%%%%%%% FIG 15 %%%%%%%%%%%%%%%%%%%%%%%%%%%%%%%%%

\begin{figure}
\epsscale{1.1}
%\plotone{cfr135_15.eps}
\plotone{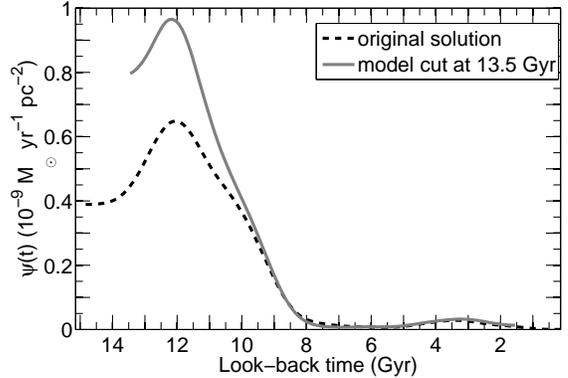}
\caption{Comparison between the Cetus $\psi(t)$ obtained 
using a model CMD with constant SFR up to 15 (dashed black line) and 
and up 13.5 Gyr (solid grey line).
\label{fig:135_15}}
\end{figure}

%%%%%%%%%%%%%%%%%%%%%%%%%%%%%%%%%%%%%%%%%%%%%%%%%%%%%%%%%%%%%%%%%%%%%%
%%%%%%%%%%%%%%%%%%%%%%%%%%%%%%%%%%%%%%%%%%%%%%%%%%%%%%%%%%%%%%%%%%%%%%
%

To further investigate the significance of deriving ages of stars older 
than 14 Gyr, we 
performed the following test. Using the derived SFH, we calculated a
synthetic CMD, from which we removed all the stars older than 13.5 Gyr. 
We simulated the observational errors and then derived the SFH of this new
CMD identically as for Cetus. Fig. \ref{fig:135} presents the input and 
recovered $\psi(t)$. The latter shows some level of star formation at 
epochs older than 13.5 Gyr, when there should be none. 
This means that the combination of observational errors and the uncertainties
associated with the SFH algorithms is responsible for the smearing  
of the main peak of star formation at the oldest epochs.

Therefore, we decided to adopt the external constraint on the age
of the Universe coming from the cosmic microwave background experiments,
and we set the upper limit to the model stars' age to 13.5 Gyr.
This choice is also supported by the fact that current stellar evolution models
give Milky Way globular cluster ages in very good agreement with this 
estimated age of the Universe \citep{marin09}.
The effect of this constraint is shown in Fig. \ref{fig:135_15}. 
The main effect of limiting the age of 
the oldest stars in the synthetic reference CMD is that the peak
of star formation is higher (since total star formation must be conserved).
However, there is no effect on the age or the end of the main episode
of star formation, nor on 
the star formation at epochs younger than 10 Gyr. With both 
models, we calculate that Cetus completely stopped forming stars $\sim$ 
8 Gyr ago.
This test was repeated with both the DAOPHOT and DOLPHOT solutions,
leading to the same conclusion. Thus, we adopt this constraint on
the SFH for the rest of our analysis.

%%%%%%%%%%%%%%%%%%%%%%%%%%%%%%%%%%%%%%%%%%%%%%%%%%%%%%%%%%%%%%%%%%%%%%
%%%%%%%%%%%%%%%%%%%%%%%%%%%%% FIG 16 %%%%%%%%%%%%%%%%%%%%%%%%%%%%%%%%%

\begin{figure}
\epsscale{.90}
%\resizebox{8truecm}{7truecm}{\includegraphics[clip=true]{cetus-sfh3d40-daophot-135.eps}}
%\resizebox{8truecm}{7truecm}{\includegraphics[clip=true]{cetus-sfh3d40-dolphot-135.eps}}
\resizebox{8truecm}{7truecm}{\includegraphics[clip=true]{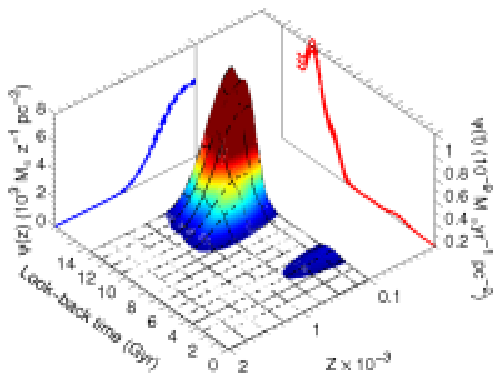}}
\resizebox{8truecm}{7truecm}{\includegraphics[clip=true]{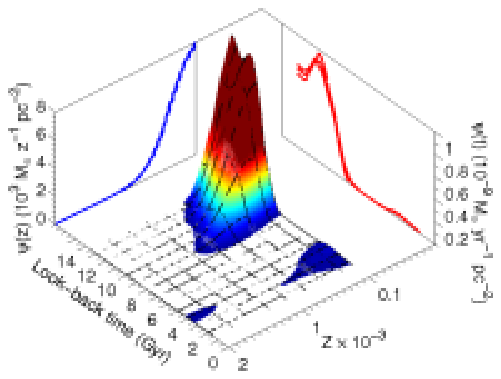}}
\caption{3-D representation of the Cetus SFH, derived with the DAOPHOT 
{\itshape (top)} and the DOLPHOT {\itshape (bottom)} photometry. Both 
show the average of 24 solutions calculated at the minimum $\chi^2_\nu$ in 
the $\delta mag$-$\delta col$ grid. The $\chi^2_{\nu, min}$  is 2.18 and 
2.92 for the DAOPHOT and DOLPHOT solutions, respectively.
\label{fig:sfhbest}}
\end{figure}

%%%%%%%%%%%%%%%%%%%%%%%%%%%%%%%%%%%%%%%%%%%%%%%%%%%%%%%%%%%%%%%%%%%%%%
%%%%%%%%%%%%%%%%%%%%%%%%%%%%%%%%%%%%%%%%%%%%%%%%%%%%%%%%%%%%%%%%%%%%%%

%
%%%%%%%%%%%%%%%%%%%%%%%%%%%%%%%%%%%%%%%%%%%%%%%%%%%%%%%%%%%%%%%%%%%%%%
%%%%%%%%%%%%%%%%%%%%%%%%%%%%% FIG 17 %%%%%%%%%%%%%%%%%%%%%%%%%%%%%%%%%

\begin{center}
\begin{figure*}
\epsscale{1.0}
%\resizebox{15truecm}{15truecm}{\includegraphics[clip=true]{cetus-135_summary.eps}}
\resizebox{15truecm}{15truecm}{\includegraphics[clip=true]{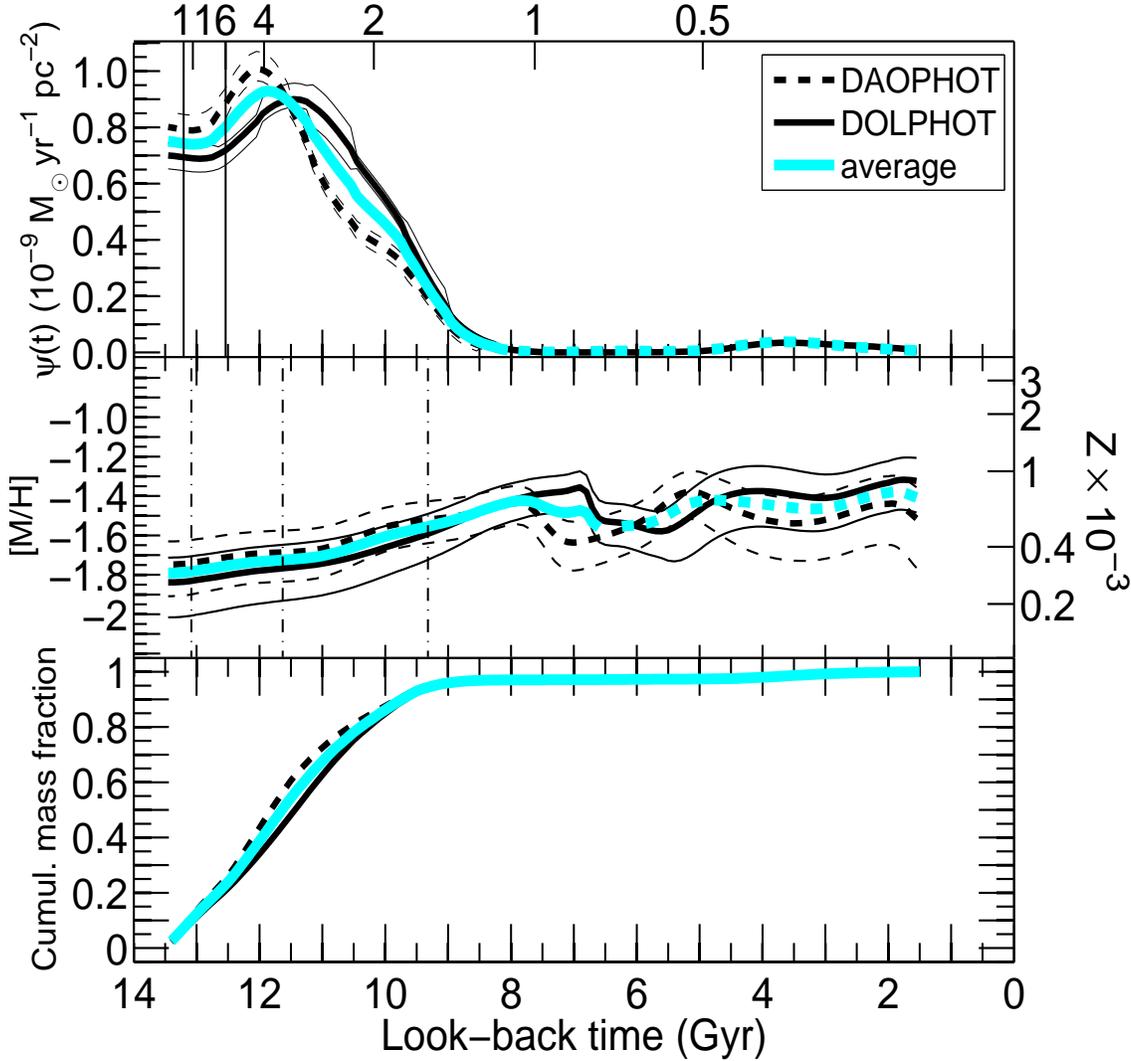}}
\caption{Summary of the Cetus SFH results. The three panels represent,
from top to bottom, the $\psi(t)$, the age-metallicity relation 
and the cumulative mass fraction. We show the results with the BaSTI 
library for both the DAOPHOT (dashed line) and the DOLPHOT (continuous)
photometry, and the average of the two (thick cyan line).  The dashed 
part of the cyan line representing the average of the two sets of 
photometry in the middle panel indicates the presence of BSs rather
than a low level of intermediate-age and young SF. The thin lines 
represent the error bars, while the vertical lines in the upper panel 
mark redshift {\itshape z} $=$ 15 and 6, that is the epochs corresponding 
to the reionization. \citep{bouwens07}. The dotted-dashed lines in the central
panel mark the epochs when 10\%, 50\% and 90\% of the mass was formed.
\label{fig:sfh_daodol}}
\end{figure*}
\end{center}

%%%%%%%%%%%%%%%%%%%%%%%%%%%%%%%%%%%%%%%%%%%%%%%%%%%%%%%%%%%%%%%%%%%%%%
%%%%%%%%%%%%%%%%%%%%%%%%%%%%%%%%%%%%%%%%%%%%%%%%%%%%%%%%%%%%%%%%%%%%%%
%

	\subsection{The final SFH of Cetus}\label{sec:sfh}

The solutions for each set of photometry, calculated as the average of the 24 
solutions at the $\chi^2_{\nu, min}$, are represented in Fig. \ref{fig:sfhbest}
where a 3-D smoothed histogram summarizes the main features. This is a 
useful representation giving an overall view of the SFH of a system 
\citep{hodge89}, including the star formation rate as a function of 
time, the metallicity distribution function and the age-metallicity 
relation. It shows graphically how the oldest stars are more metal-poor, 
and how chemical enrichment proceeds toward younger ages. The two projections
$\psi(z)$ and $\psi(t)$ are the metallicity distribution integrated over 
time (blue line) and the age distribution integrated over metallicity 
(red line). {Errors have been calculated for $\psi(t)$ and $\psi(z)$ as the
dispersion of the 24 individual solutions used in the average 
\citep[see ][]{iacpop}. } In particular, the $\psi(z)$, shows a peak at the most metal 
poor regime, and a sharp decrease toward the more metal-rich tail. Only 
a negligible fraction of stars ever formed have metallicities higher 
than Z=10$^{-3}$. 

The overall agreement between the DAOPHOT and the DOLPHOT solutions, 
illustrated in Fig. \ref{fig:allbest_daodol} and in Fig. \ref{fig:sfhbest} 
suggests that we can safely adopt the average of the two solutions for 
the SFH of the Cetus dSph galaxy,  and the difference of the two as an
indication of external errors \footnotemark[24].

\footnotetext[24]{From Fig. \ref{fig:external}, middle panel, it can be concluded
that the adopted photometry has a larger influence on the SFH than the stellar
evolution models used. The same is true with respect to assumptions on binaries
(see Fig. \ref{fig:sfr4680}, top panel), or the IMF (Skillman et al. in preparation).
Therefore, using the difference in the SFH calculated from the two photometry
sets as an indication of external errors seems justified.}

Fig. \ref{fig:sfh_daodol} summarizes the main results by comparing the 
star formation rate as a function of time (top), the age-metallicity relation
(middle), and the cumulative mass function (bottom), for both sets of photometry 
with the adopted average calculated as explained in Hidalgo et al., in prep. %\citet{hidalgolgs3}.
The thick dashed and continuous lines represent the DAOPHOT and DOLPHOT 
solutions, while the thin lines are the corresponding error bars, calculated
as the 1-$\sigma$ dispersion of the 24 solutions at $\chi^2_{\nu, min}$.
This is a statistically meaningful way of defining the uncertainities introduced 
by the SFH recovery method, as extensively discussed in \citet{iacpop}.
The thick cyan line is the average of the solutions obtained for the two photometries.
Fig. \ref{fig:sfhbest} and Fig. \ref{fig:sfh_daodol} provide two complementary
ways of presenting the derived SFH of Cetus. From these figures we see that
Cetus is mostly an old, metal-poor stellar system, with the vast majority 
of stars older than 8 Gyr and more metal-poor than Z = 0.001.
Tab. \ref{tab:tab1} summarizes the main integrated quantities derived for both 
photometry sets.
%Tab. \ref{tab:tab2} summarizes the build up of mass in Cetus as a function of time.

%%%%%%%%%%%%%%%%%%%%%%%%%%%%%%%%%%%%%%%%%%%%%%%%%%%%%%%%%%%%%%%%%%%%%%%%%%%%%%%%
%%%%%%%%%%%%%%%%%%%%%%%%%%%%%%%%%%%%%%%%%%%%%%%%%%%%%%%%%%%%%%%%%%%%%%%%%%%%%%%%

\begin{deluxetable}{lcc}
\tabletypesize{\scriptsize}
\tablewidth{0pt}
\tablecaption{Integrated quantities derived for the Cetus dSph.\label{tab:tab1}}
\tablehead{
\colhead{ } & \colhead{DAOPHOT} & \colhead{DOLPHOT}}
\startdata
$\int\psi(t)dt$ $[10^{6}M_{\odot}]$                    &  (1.81$\pm$0.05)  &  (1.91$\pm$0.07) \\
$<\psi(t)>       [10^{-7} M_{\odot} yr^{-1} pc^{-2}$]  &  (1.17$\pm$0.02)  &  (1.23$\pm$0.03) \\
$<age>           [10^{10}yr]$                          &  (1.14$\pm$0.02)  &  (1.12$\pm$0.02) \\
$<[Fe/H]>        [10^{-4} dex]$                        &  (4.23$\pm$0.54)  &  (3.64$\pm$0.53) 
\enddata
%\tablenotetext{a}{}
\end{deluxetable}

%%%%%%%%%%%%%%%%%%%%%%%%%%%%%%%%%%%%%%%%%%%%%%%%%%%%%%%%%%%%%%%%%%%%%%%%%%%%%%%%
%%%%%%%%%%%%%%%%%%%%%%%%%%%%%%%%%%%%%%%%%%%%%%%%%%%%%%%%%%%%%%%%%%%%%%%%%%%%%%%%

The vertical lines in the top panel indicate the redshift $z$ = 15 and $z$ = 6, which 
mark the epoch when reionization was under way, the latter considered to be
the epoch when the Universe was fully reionized \citep[][ and references therein]{bouwens07}. 
The vertical lines in the middle panel of Fig. \ref{fig:sfh_daodol} mark 
the epoch when 10, 50 and 90\% of the stellar mass was formed. They indicate that
Cetus experienced the bulk of its star formation before redshift $\approx 2$,
with a peak at $z \sim 4$.

In the following, we focus on two important aspects of the Cetus SFH: the sequence
of candidate BSs and the actual duration of the main episode of star formation.

		\subsubsection{Recent star formation or blue stragglers}\label{sec:bs}

\begin{figure}
\epsscale{1.2}
%\plotone{cmdbs.eps}
\plotone{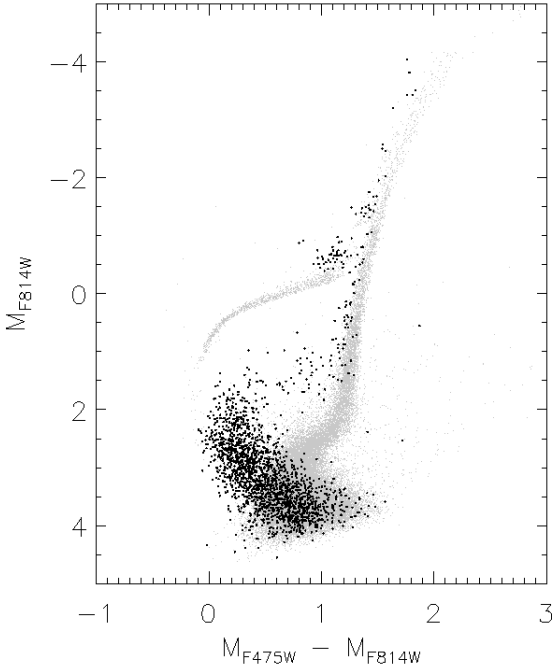}
\caption{Syntetic CMD corresponding to the SFH of Cetus. The young metal-poor
stars are highlighted with bigger points. Their position on the CMD is fully 
consistent with the candidate BSs sequence.
\label{fig:cmdbs}}
\end{figure}

In \S \ref{sec:cmd} we gave circumstantial evidence, based on the
comparison with isochrones, that the objects bluer and brighter than the
old MSTO are BSs, rather than truly young stars. More evidence of this 
comes from the analysis of the SFH presented in \ref{fig:sfhbest}.
The plot shows an event of very low SFR that produced a small percentage of
metal-poor but relatively young ($\sim$ 2-4 Gyr old) stars. This population
clearly does not follow the general age-metallicity relation. 
Fig. \ref{fig:cmdbs} shows that the position of these stars in the solution CMD
is fully consistent with the position of the blue plume on the CMD. This
feature is systematically present in all our solutions, and does not
depend on the sampling, library, or photometry adopted. Given the
age-metallicity relation derived for Cetus, there is no reason to expect a
relatively young but very metal-poor population - this would require something
like an episode of late infall of very metal poor gas. This interpretation is
nicely  supported by the age-metallicity relation presented in Fig.
\ref{fig:sfh_daodol} (middle panel), which shows a continuous increase of
metallicity with time, until a peak occurs 8 Gyr ago. After that, we observe a
sharp decline followed by a renewed increase. If we accept that blue
stragglers are found in all or nearly all sufficiently populous open and
globular star clusters as well as in the Milky Way field halo population, then there
is no conflict  between our data and the assertion that no star formation
occurred in Cetus in the last 8 Gyr; the most natural explanation for these
stars is that they are blue stragglers belonging to the old, metal-poor
population. Note that the contribution of BS to the total estimated
mass of Cetus is $\leq 3\%$. Therefore, the presence of fainter BSs 
contaminating the TO and MS regions is expected to have negligble impact
on the derived SFH.
An in-depth discussion of the candidate BSs in both dSph
galaxies of the LCID sample will be presented in a forthcoming paper
(Monelli et al., in prep.)

%%%%%%%%%%%%%%%%%%%%%%%%%%%%%%%%%%%%%%%%%%%%%%%%%%%%%%%%%%%%%%%%%%%%%%
%%%%%%%%%%%%%%%%%%%%%%%%%%%%% FIG 18 %%%%%%%%%%%%%%%%%%%%%%%%%%%%%%%%%

%%%%%%%%%%%%%%%%%%%%%%%%%%%%%%%%%%%%%%%%%%%%%%%%%%%%%%%%%%%%%%%%%%%%%%
%%%%%%%%%%%%%%%%%%%%%%%%%%%%%%%%%%%%%%%%%%%%%%%%%%%%%%%%%%%%%%%%%%%%%%

		\subsubsection{The short first episode of SFH in Cetus} \label{sec:short}

The description presented in \S  \ref{sec:results} corresponds to the 
SFH of Cetus as derived using the discussed algorithms. It is clear that 
this solution is the result of the convolution of the {\itshape actual} 
Cetus SFH with some smearing produced by the observational errors, and the 
limitations of the SFH recovery method \citep{iacpop}. In the following, 
we will try to further constrain the features of the actual Cetus SFH 
through some tests with mock stellar populations.

To investigate the actual duration of the dominant episode of star formation,
and to assess the age resolution at the oldest ages, we performed the following 
test. We created mock galaxies characterized by fixed metallicity and short 
episodes of star formation around the mean value of 12.0 Gyr (see Fig. 
\ref{fig:burst}). Since a Gaussian profile fit to the Cetus $\psi(t)$ 
yields an estimate of $\sigma = 1.53$ Gyr, we used three $\psi(t)$ modeled
by a Gaussian profile with $\sigma = 0.5$, $1.0$, and 1.5 Gyr.
We simulated the observational errors in the mock stellar populations, 
and recovered their SFH using the same prescriptions as for the real data.
For simplicity, only the completeness tests of the DAOPHOT photometry 
were adopted to simulate the observational errors in the mock CMD.
The number of stars in the mock populations was such that the total number of
stars in the bundles was comparable to the case of the real galaxy.

The first important result of this test is that the age of the peak 
of the recovered mock population is systematically well recovered, within
0.2 Gyr. Thus, if the initial episode of star formation in Cetus is 
well represented by a Gaussian profile, then we are confident in our
ability to determine the time of the peak star formation rate.
Similar tests have been also discussed
in \citet{iacpop} and Hidalgo et al., in prep. %\citet{hidalgolgs3}.
The second finding is that the duration of the recovered $\psi(t)$ is 
wider than the input one. In particular, we measured the HWHM and 
calculated the corresponding $\sigma$, which are equal to 1.44, 1.61, 
and 1.95 Gyr for the three episodes of increasing duration. 
The increase from the input $\sigma$ is a consequence of the
blurring effect of the observational errors and the limited age
resolution at old ages.

The comparison of the duration of the main peak in the Cetus and 
in the mock stellar populations suggests that the estimated duration 
for Cetus is an {\itshape upper limit}. Moreover, we note that this
value can be further constrained by the $\sigma = 0.5$ 
and the $\sigma = 1.0 $ Gyr Gaussian profile tests, with a result 
close to $\sigma = 0.8$ Gyr (interpolated value). 
Moreover, we can expect that the limited age resolution
broadens both the older and the younger sides of the $\psi(t)$, around its
maximum. This is supported by Fig. \ref{fig:burst}, where we can compare
the youngest epoch of star formation in the three models and in their
recovered profiles. The comparison reveals that the output $\psi(t)$
is more extended to younger ages than the input ones, and that this 
effect is less important for increasing $\sigma_{in}$. This suggests 
that at least part of the star formation between 10 and 8 Gyr ago 
derived in Cetus is not real. However, we stress that the differences 
between the actual SFH and the recovered one affect only 
the duration of the star formation solution and the height of the main 
peak, but they do not alter significantly the age of the peak itself.

%
%%%%%%%%%%%%%%%%%%%%%%%%%%%%%%%%%%%%%%%%%%%%%%%%%%%%%%%%%%%%%%%%%%%%%%
%%%%%%%%%%%%%%%%%%%%%%%%%%%%% FIG 19 %%%%%%%%%%%%%%%%%%%%%%%%%%%%%%%%%

\begin{figure}
\epsscale{1.2}
%\plotone{cetus_burst_summary_areanormalized.eps}
\plotone{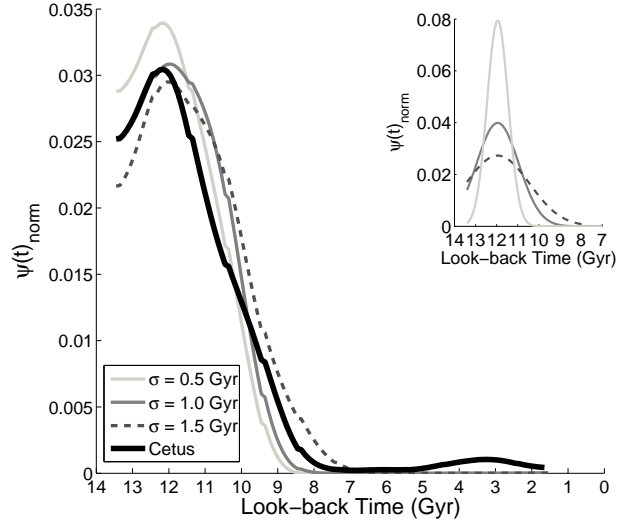}
\caption{Comparison of the $\psi(t)$ derived for Cetus with that
retrieved from three mock stellar populations. These were computed with a 
Gaussian profile $\psi(t)$ with $\sigma = $0.5, 1.0, and 1.5 Gyr peaked at 12 Gyr,
as shown in the small top-right panel. The recovered $\psi(t)$ is wider than the
input one. This test indicates that the intrinsic duration of the main 
star formation episode in Cetus can be confined between the $\sigma = $0.5
and $\sigma = $1 Gyr. All the curves presented are normalized to their total area.
\label{fig:burst}
\label{fig:burst}}
\end{figure}

%%%%%%%%%%%%%%%%%%%%%%%%%%%%%%%%%%%%%%%%%%%%%%%%%%%%%%%%%%%%%%%%%%%%%%
%%%%%%%%%%%%%%%%%%%%%%%%%%%%%%%%%%%%%%%%%%%%%%%%%%%%%%%%%%%%%%%%%%%%%%
%

\begin{figure*}
\epsscale{1.}
%\plotone{cmdbest_daophot.eps}
\plotone{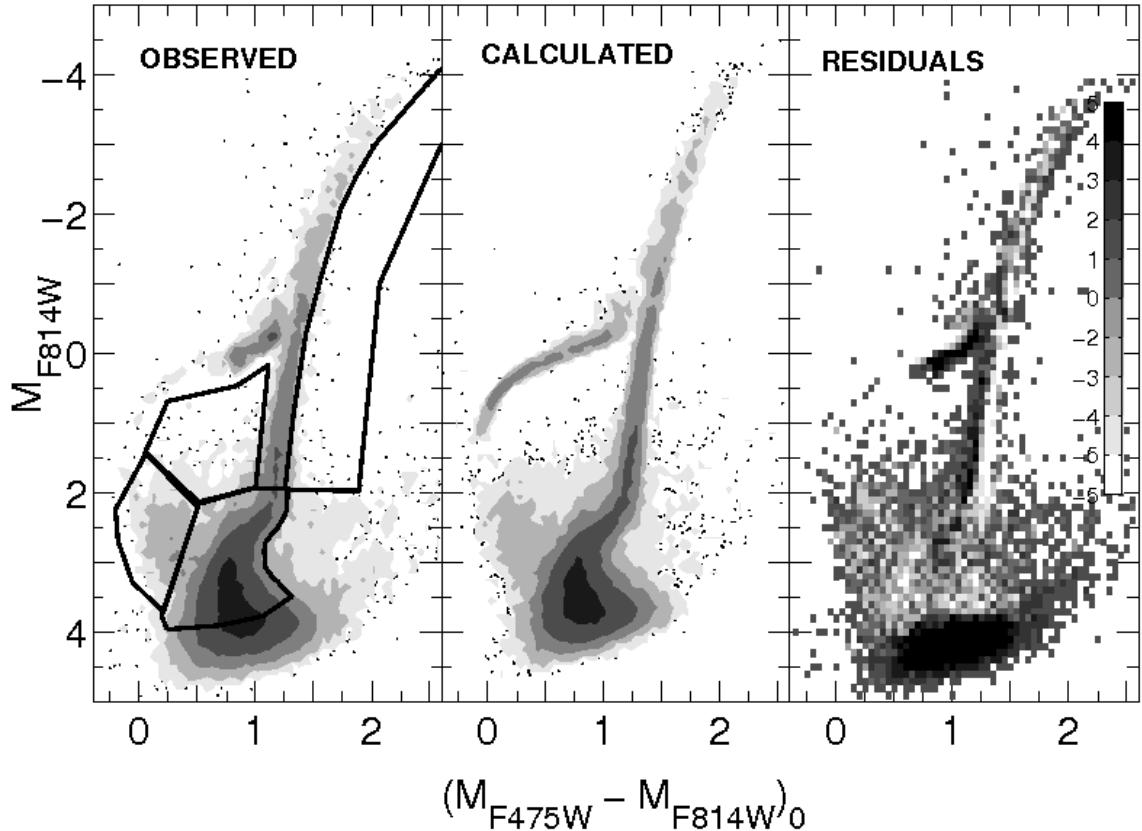}
\caption{Comparison of the observed (left) and solution (centre) CMDs.
The latter was derived extracting random stars from the model CMD
used to derive the solution, in such a way that each simple population
contributes proportionally to the calculated SFR.
The rigth panels shows the residual Hess diagram, shown in unit of
Poisson errors. The Figure disclose an overall general agreement
between the observed and the solution CMDs, particularly in the whole TO
region used to calculate the SFH. The discrepancy at the faintest
magnitude is due to the limit magnitude of the model CMD. Note
that such discrepancy does not affect the portion of the CMD included
in the bundles, which are repeated in the left panel. The differences
in the HB morphology are expected, since small differences in the 
adopted mass-loss along the RGB have a big impact on the HB morphology. 
\label{fig:cmdbest}}
\end{figure*}

Finally, Fig. \ref{fig:cmdbest} presents a comparison between the observed 
DAOPHOT CMD and a synthetic CMD corresponding to the best solution (left 
and cetral panels). We find a satisfactory agreement in all the main evolutionary 
phases, including the RGB phase which has not been used to derive the solution.
As far as it concerns the differences in the HB morphology, it is worth 
noting that we have not tried to properly reproduce this observational 
feature. Due to the strong dependence of the HB morphology on the mass 
loss efficiency along the RGB, it is clear that a better agreement between 
the observational data and the synthetic CMD could be obtained with a
better knowledge of mass loss efficiency and its spread.
The right panel shows the residual Hess diagram, which supports the good
agreement between the observed and the solution CMD.
%
%%%%%%%%%%%%%%%%%%%%%%%%%%%%%%%%%%%%%%%%%%%%%%%%%%%%%%%%%%%%%%%%%%%%%%
%%%%%%%%%%%%%%%%%%%%%%%%%%%%% FIG 20 %%%%%%%%%%%%%%%%%%%%%%%%%%%%%%%%%

%%%%%%%%%%%%%%%%%%%%%%%%%%%%%%%%%%%%%%%%%%%%%%%%%%%%%%%%%%%%%%%%%%%%%%
%%%%%%%%%%%%%%%%%%%%%%%%%%%%%%%%%%%%%%%%%%%%%%%%%%%%%%%%%%%%%%%%%%%%%%
%

	\subsection{Are we seeing the effects of reionization in the Cetus SFH?}\label{sec:reion}

One of the main drivers of the LCID project is to find out whether cosmic
reionization left a measurable imprint on the SFHs of nearby, isolated
dwarf galaxies. The Universe is thought to have been completely reionized
before {\itshape z } $\geq$ 6 \citep{bouwens07}, corresponding to T
$\simeq$ 12.8 Gyr ago. The comparison of these values
with the Cetus $\psi(t)$ in Fig. \ref{fig:sfh_daodol} (see upper panel),
shows that the vast majority of the {\itshape measured} star formation in Cetus
occurred at  epochs more recent than 12.8 Gyr ago (see also Tab.
\ref{tab:tab2}). This is also true when the {\itshape actual} Cetus SFH is
considered, as discussed in section \ref{sec:short}. Therefore, we  conclude that
there is no evident coincidence between the epoch when reionization was
complete and the time in which the star formation in Cetus ended.

\begin{figure*}
\epsscale{1.}
%\plotone{cmdbest_daophot.eps}
\plotone{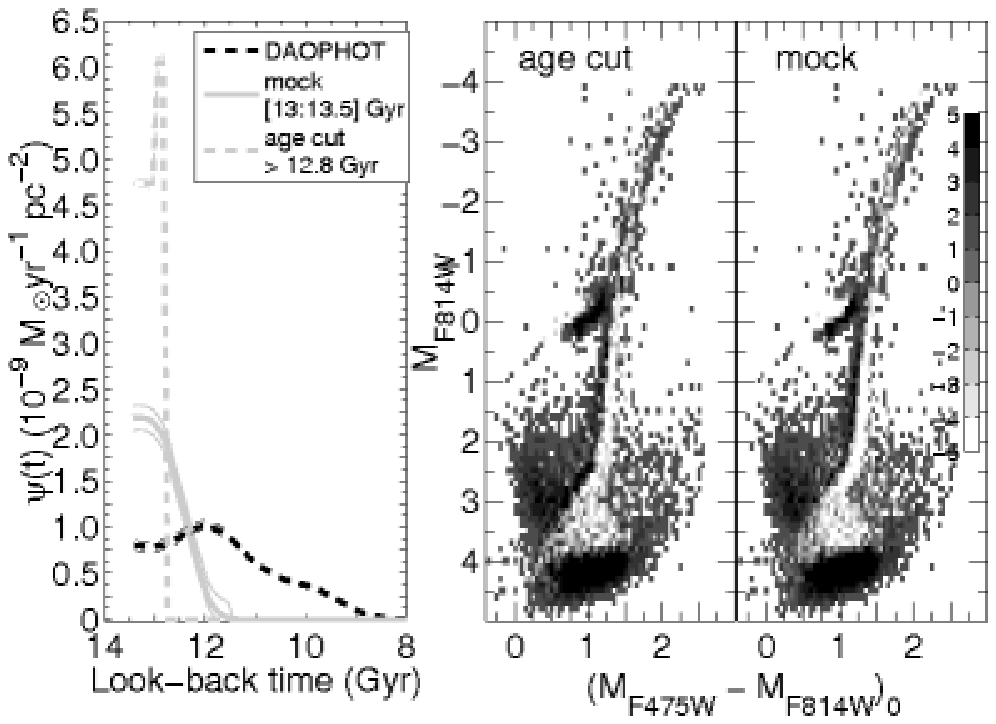}
\caption{The left panel compares the $\psi(t)$ of Cetus (black dasked line)
to (i) the one of a mock population, created with stars in the range between
13 and 13.5 Gyr (solid grey line), and (ii) a solution derived for Cetus, 
but imposing the constraint to use only stars older the 12.8 Gyr (dashed grey line).
The central and right panels show the residual Hess diagram of these two
solutions, calculated identically as in Fig. \ref{fig:cmdbest}. The 
residual CMDs show important systematics suggesting that these solutions
are not compatible with the derived best solution of Cetus, thus supporting
that no strong correlation exists between the epoch of the reionization 
and the epoch of the star formation in this galaxy.
\label{fig:reion}}
\end{figure*}

However, one might ask the question of whether acceptable solutions could
be found that don't show star formation more recent than 12.8 Gyr ago,
i.e., that are compatible with the hypothesis that reionization contributed
to the supression of star formation in Cetus. With this question in mind,
we performed two additional tests, summarized in Fig. \ref{fig:reion}.

$\bullet$ First, we calculated a solution SFH of Cetus costraining the 
model CMD to stars of ages older than 12.8 Gyr. This is not the typical 
approach for the IAC method, but nontheless the  code looks for the
absolute minimum in the allowed parameter space, and provides the best
possible solution (grey dashed line). However, we found that this solution
is clearly at odds with the derived best SFH of Cetus, and not compatible
with it. Also, note the significantly higher $\chi^2_\nu$, which increases 
to 7.80, and the large residuals (Fig. \ref{fig:reion}, central panel). The 
latter, calculated with respect to the observed CMD identically as in Fig.
\ref{fig:cmdbest}, indicate important systematics in the solution. The 
excess of stars in the solution CMD in the red part of the TO (lightest bins),
indicates the presence of too many stars in this region. This is in
agreement with the overall older population, with the main peak at $\sim$
13 Gyr ago.

$\bullet$ We recovered the SFH of a mock population, similarly to the tests
shown in Fig. \ref{fig:burst}, created with constant SFR in the age range
between 13 and 13.5 Gyr, and zero elsewhere. The solution shown in Fig.
\ref{fig:reion}, discloses that the peak is recovered at the correct age,
in the oldest possible age bin (grey solid line). This suggests that an
actual SFH where most of stars were born before the end of the reionization 
epoch is not compatible with the solution we derived for Cetus. Moreover, 
the general trend in the residual plot (Fig. \ref{fig:reion}, right panel), 
is similar to  the previous case, with an excess of red TO stars.

These two tests, together with the ones shown in Fig. \ref{fig:burst},
suggest that, on the basis of the available data coupled with the SFH 
recovery code, we can reject the hypothesis that the majority of the
star formation occurred in Cetus in epochs older than {\itshape z}
$\simeq 6$.

%

%
%%%%%%%%%%%%%%%%%%%%%%%%%%%%%%%%%%%%%%%%%%%%%%%%%%%%%%%%%%%%%%%%%%%%%%
%%%%%%%%%%%%%%%%%%%%%%%%%%%%% FIG 21 %%%%%%%%%%%%%%%%%%%%%%%%%%%%%%%%%

%%%%%%%%%%%%%%%%%%%%%%%%%%%%%%%%%%%%%%%%%%%%%%%%%%%%%%%%%%%%%%%%%%%%%%
%%%%%%%%%%%%%%%%%%%%%%%%%%%%%%%%%%%%%%%%%%%%%%%%%%%%%%%%%%%%%%%%%%%%%%
%

	\subsection{Summary}\label{sec:summa_sfh}

The various pieces of information analysed in the \S \ref{sec:results}
can be sumarized as follows.
The star formation of the observed area of Cetus shows a well-defined peak
that occurred $\approx$ 12 Gyr ago. By resolving the star formation of
mock stellar populations, we showed that the actual duration of the main event
is probably overestimated, due to the smearing effect of the observational
errors. This does not affect the age of the main peak, but only its
duration, at both older and younger epochs. Therefore, the last epoch of
star formation, estimated at 8 Gyr ago, is probably actually somewhat older.
The various evidence indicating that the blue plume of objects identified
in the CMD consists of a population of BS, supports the idea that 
no star formation occurred in Cetus in the last 8-9 Gyr.
In Tab. \ref{tab:tab2} we present the epoch when various fractions of the
total mass were formed, once we excluded the contribution of the BS
population. However, the actual values of the percentiles reported could 
be slighlty different due to the smearing of the solution with 
respect of the actual, underlying SFH (see \S \ref{sec:short}).

The derived SFH is at odds with the conclusion by \citet{sarajedini02}, 
who suggested that Cetus might be 2-3 Gyr younger that the old
Galactic globular clusters. The stability of the main peak around the age
of 12 Gyr, together with up-to-date estimates of the globular cluster ages
\citep{marin09}, suggest that such a difference can be ruled out.

%%%%%%%%%%%%%%%%%%%%%%%%%%%%%%%%%%%%%%%%%%%%%%%%%%%%%%%%%%%%%%%%%%%%%%%%%%%%%%%%

\begin{deluxetable}{ccc}
\tabletypesize{\scriptsize}
\tablewidth{0pt}
\tablecaption{Mass percentiles formed in Cetus as a function
of redshift and look-back time.\label{tab:tab2}}
\tablehead{
\colhead{Mass \%} & \colhead{redshift} & \colhead{Look-Back Time (Gyr)} }
\startdata
  10   &   8.8  &   13.1   \\
  20   &   5.7  &   12.7   \\
  30   &   4.5  &   12.3   \\
  40   &   3.7  &   12.0   \\
  50   &   3.2  &   11.6   \\
  60   &   2.8  &   11.3   \\
  70   &   2.4  &   10.9   \\
  80   &   2.0  &   10.4   \\
  90   &   1.7  &    9.7   \\

\enddata
%\tablenotetext{a}{}
\end{deluxetable}

%%%%%%%%%%%%%%%%%%%%%%%%%%%%%%%%%%%%%%%%%%%%%%%%%%%%%%%%%%%%%%%%%%%%%%%%%%%%%%%%

\section{Discussion}\label{sec:discu}

The SFH of Cetus derived in this paper shows that this dSph was able to form the vast 
majority of its stars coincident with and after the epoch of reionization. 
Furthermore, there are no signs in the SFH suggestive of a link between a 
characteristic time in the SFH and the time of reionization. This may be due 
to the fact that Cetus, with a velocity dispersion of $\sigma = $ 17 $\pm$ 
2 km/s \citep{lewis07}, is one of the brightest ($M_V\sim -10.1$, 
\citealt{whiting99}), and presumably 
one of the most massive dSphs in the Local Group. Its large mass might therefore 
have shielded its baryonic content from the effects of reionization 
\citep{barkana99, carraro01, tassis03, ricotti05, gnedin06, okamoto09}. The 
latter reference, in particular, argues that all classical LG dwarfs originate 
in the few surviving satellites which, at the time of reionization, had potential 
wells deeper than the threshold value of $\sim 12$ km/s.  
If reionization by the UV background cannot be related to the early
truncation of star formation in Cetus, then other circumstances must be at
play. According to the model by \citet{okamoto09}, the 
sudden truncation in the SFH of dwarfs like Cetus arises from having their gas 
reservoirs stripped during first infall into the potential wells of one of the 
massive LG spirals (M31 or MW). Another possibility frequently raised in the 
literature is that the end of star formation reflects the blow-out of gas driven 
by supernova explosions. We examine each of these two possibilities below.

The radial velocity of Cetus measured by \citet{lewis07} implies that this galaxy 
is receding with respect to the Local Group barycenter and, under reasonable 
assumptions about its proper motion and the mass distribution in the Local Group, 
that it is close to the apocenter of its orbit.  Therefore, it is possible that 
Cetus is in a highly radial orbit and that it has gone through pericenter at least 
once. If this was the case, then it is possible that the truncation of the star 
formation occurred due to the effect of tides and ram-pressure \citep{mayer01, okamoto09} 
during pericentric passage. The mild evidence of rotation found by \citet{lewis07} 
may be interpreted to imply that the progenitor of Cetus was a disk-like dwarf that 
underwent an incomplete transformation into a spheroidal system as a result of the 
strong tides operating at pericenter \citep{mayer01, lokas09}. The models by 
\citet{lokas09} predict different final configurations after one passage for a 
disky dwarf. Depending on the geometry of the encounter, either a triaxial, 
rotating system or a prolate object with little rotation may result. 
Alternatively, it has been proposed by
\citet{donghia09} that a gravitational resonance mechanism, coupling the
rotation of the accreting disky dwarf with its orbital motion, is able to turn 
it into a dSph in a single encounter. However, this requires a nearly prograde
encounter. The latter condition may not be common during the hierarchical 
assembly of the galaxy, but may still have occurred in just a few cases 
(prograde accretions are seen in cosmological hydrodynamical simulations)
which would be enough to explain the small subset of distant dSphs. 
A firm measure of the shape of Cetus might thus be an important step in the context 
of constraining its formation models. 
Independently of the mechanism driving the encounter, if Cetus has indeed 
been through the pericenter of its orbit, it is possible that it owes its 
present-day isolation in the LG to the fact that it gained orbital 
energy in the process and is now on a weakly bound orbit that takes it well beyond 
the normal confines of the LG. This possibility has been highlighted by \citet{sales07} 
and \citet{ludlow09}, who discuss the possibility that the odd orbital motion of Cetus 
might have originated in a multiple-body interaction during pericenter.

On the other hand, models of the SFH in dwarfs driven by feedback are also able to 
reproduce the main properties of dSph galaxies without invoking galaxy interactions.
In particular, \citet{carraro01} find that the key parameter is the dwarf's central 
density, while \citet{sawala09} argue that, as in previous models such as \citet{maclow99}, 
low-mass systems are those most likely to end up as dSph galaxies. However, despite some 
differences all models of gas blow-out by supernova explosions find the mechanism to 
be inefficient above $10^7 M_{\odot}$.  While ultra-faint dwarfs may well be in the 
right mass regime, this is a problem for all classic dSphs since their initial 
masses were likely at least two orders of magnitude above this threshold \citep{mayer09}.
Note also, that models of dSph formation that are based only on galaxy mass and
feedback are unable to reproduce the morphology-density relationship.

When compared to the dSph satellites of the Milky Way, the stellar content of Cetus 
shows striking similarities with the Draco dSph. \citet{aparicio01}, on the basis 
of a ground-based CMD of depth similar to those obtained by LCID for more distant 
dwarfs, concluded that most of its star formation occurred more than 10 Gyr ago, 
and that Draco stopped forming stars $\sim$ 8 Gyr ago. They also infer a less 
intense episode $\sim$3-2 Gyr ago, which is reminiscent of the population in Cetus 
that we associate with blue stragglers. Therefore, Cetus's SFH  looks more similar
to that of other nearby dSphs with mostly old stellar populations, such as Draco, 
Ursa Minor, Leo I, Sculptor or Sextans. Their SFHs differ from those of other dSphs 
like Carina, Fornax or Leo II, where there is clear evidence for a substantial 
intermediate-age population. The striking observation that Cetus and Tucana
(the other isolated LG dSph) do not follow the morphology-density relation 
of Milky Way dSph galaxies observed in the LG today appears to be simply
an observational coincidence.  Their radial velocities indicate that it is
possible that they were much closer to the dominant LG spirals when their
SF ceased, indicating that their star formation might have been
truncated by the same mechanism responsible for the formation of the 
other LG dSphs.

\section{Summary and conclusions}\label{sec:summa}

We used HST/ACS observations to derive the SFH of the isolated Cetus 
dSph. This effort is part of the LCID project, whose main goal is to 
investigate the SFH of isolated dwarf galaxies in the LG. 
To asses the robustness of our conclusions, the LCID methodology uses and 
compares the results of two photometric codes, two stellar evolution 
libraries, and three SFH-estimation algorithms. This allowed us to check the
stability of our solution, and to give a robust estimate of the external
errors affecting it (see also Hidalgo et al., in prep.). %\citep[see also][]{hidalgolgs3}. 
The use of different stellar evolution libraries and/or SFH codes does 
not change the global picture we derived. We found that the BaSTI library
gives solutions systematically older (less than 1 Gyr) than the 
Padova/Girardi one. 
%Concerning the different SFH code, a direct comparison 
%between IAC-pop with the MATCH methods is not straightforward. IAC-pop
%benefits of the {\it dithering} approach of MinnIAC, therefore smoothing
%out the binning selection effects, both in term of simple populations (ages
%and metallicity) and boxes on the CMD, providing a robust  average SFH. MinnIAC
%also allows to easily explore a wide space of parameters, adopting
%different combinations of small color and magnitude shifts of the observed
%CMD, that were found to be very useful to compensate for eventual subtle
%calibration offsets. The MATCH method offers a more limited capability to
%explore these aspects, but nonetheless 
The cumulative mass fraction showed
in Fig. \ref{fig:external}, comparing the different 
%one individual MATCH solution with the average of 24 IAC-pop 
solutions shows a good general agreement within
the error bars. Finally, the good consistency of the SFH solutions derived 
with the DAOPHOT and DOLPHOT photometry sets warranted to adopt the average
of the two solutions as the final Cetus SFH.

Our main result is that Cetus is an old and metal-poor system. We find 
evidence for Cetus stars as old as those in the oldest populations in the 
Milky Way, and comparable in age to the age of the Universe inferred from 
WMAP \citep{bennett03}. Star formation decreased after the earliest
episode which likely peaked about $\sim$ 12 Gyr ago. We find no convincing 
evidence for a population of stars younger than $\sim 8$ Gyr old.

A number of tests aimed at recovering the SFH of mock stellar populations
%show the reliability of the MinnIAC/IAC-pop method to precisely derive
%the age of the peak of star formation, even at ages $>$ 10 Gyr, with a
yield a estimate of our precision for determining the peak of 
star formation of the order of $\leqslant$ 0.5 Gyr for a given
stellar evolution library. Therefore, our estimate that the Cetus peak 
$\psi(t)$ occurred $\sim$ 12 Gyr ago, is a robust result. Our tests also
disclosed that the measured duration of the star formation is artificially 
increased by the observational errors. We could  estimate that the intrinsic
extent is compatible with an episode approximated by a Gaussian profile
with $\sigma =$ 0.8 Gyr (or 1.9 Gyr FWHM), as opposed to the measured 
$\sigma = 1.53 $ Gyr. This likely moves the end of the star formation at older epochs. 
%Therefore, the 
%ages of the mass percentiles listed in Tab. \ref{tab:tab2} must be 
%taken as a lower limits.

%%%%%%%%%%%%%%%%%%%%%%%%%%%%%%%%%%%%%%%%%%%%%%%%%%%%%%%%%%%%%%%%%%%%%%%%%%%%%%%%

%\input{tab02}

%%%%%%%%%%%%%%%%%%%%%%%%%%%%%%%%%%%%%%%%%%%%%%%%%%%%%%%%%%%%%%%%%%%%%%%%%%%%%%%%

The age-metallicity relation shows a mild but steady increase in metallicity 
with time, up to $Z = 0.001$. The presence of bright, blue stars in the upper 
main sequence formally implies a sprinkling of stars younger than $\sim 8$ Gyr
old. These stars, however, are inferred to have metallicities below $Z=0.001$,
and thus would be clear outliers in the monotonic age-metallicity relation. 
An alternative interpretation, which we favor, is that these are not 
younger main-sequence stars but rather a population of blue stragglers with the 
same typical metallicity as the bulk of the population in Cetus.

In conclusion, we do not find any direct correlation between the epoch
of the reionization and the epoch of the star formation in Cetus.
In particular, the reionization does not seem to have been an efficient
mechanism to stop the star formation in this galaxy. Alternative mechanisms,
such as gas stripping and tidal interactions due to a close encounter
in the innermost regions of the LG, seem favored with respect of
internal processes like supernovae feedback.

%%%%%%%%%%%%%%%%%%%%%%%%%%%%%%%%%%%%%%%%%%%%%%%%%%%%%%%%%%%%%%%%%%%%%%
%%%%%%%%%%%%%%%%%%%%%%%%%%%%%%%%%%%%%%%%%%%%%%%%%%%%%%%%%%%%%%%%%%%%%%

\acknowledgments

Support for this work was provided by NASA through grant GO-10515
from the Space Telescope Science Institute, which is operated by
AURA, Inc., under NASA contract NAS5-26555, the IAC (grant 310394), the
Education and Science Ministry of Spain (grants AYA2004-06343 and
AYA2007-3E3507).
This research has made use of NASA's Astrophysics Data System
Bibliographic Services and the NASA/IPAC Extragalactic Database
(NED), which is operated by the Jet Propulsion Laboratory, California
Institute of Technology, under contract with the National Aeronautics
and Space Administration.

{\it Facilities:} \facility{HST (ACS)}.

\appendix

	\subsection{The effect of binary stars} \label{sec:bin}

\begin{figure}
\epsscale{1.0}
%\plotone{chi2new.eps}
\plotone{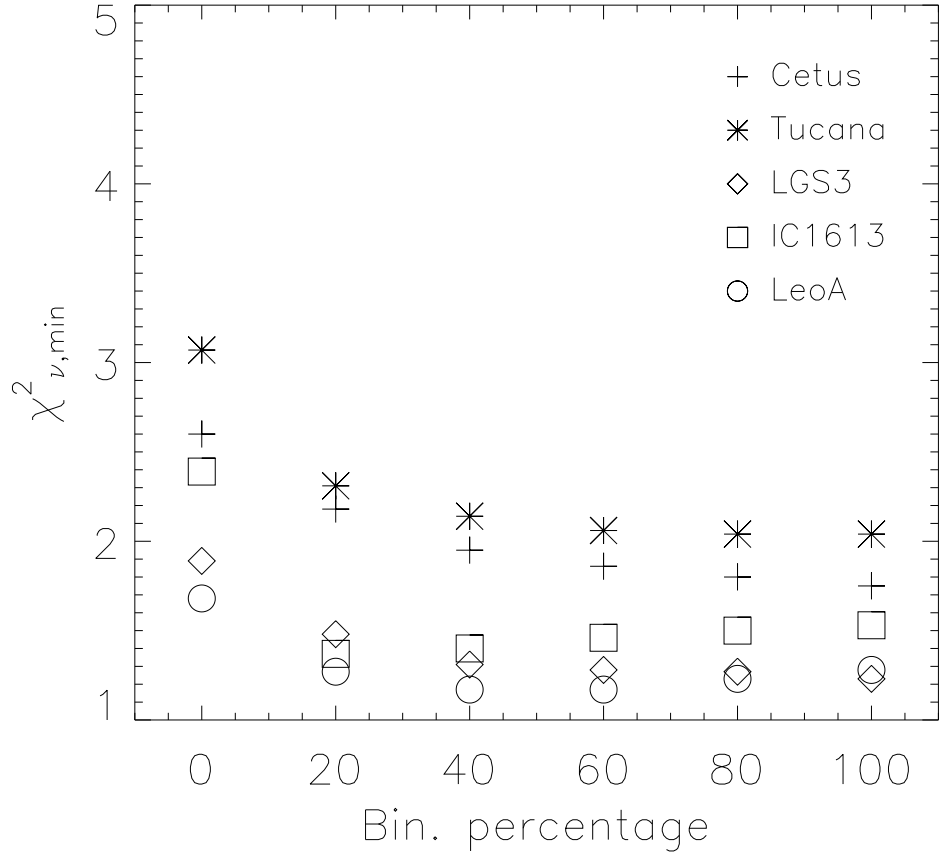}
\caption{Summary of the $\chi^2_\nu$ trend as a function of the binary fraction 
for 5 galaxies analyzed in the LCID project. 
\label{fig:chi2summary}}
\end{figure}

To study the impact of different assumptions on the binary content of the 
model CMD, we used the IAC method to the identically derive the SFH
with six different synthetic CMDs, varying the amount of binary stars from
0\% to 100\%, in steps of 20\%. In  all cases, we adopted a mass ratio
between the two components $q > 0.5$,  and a flat mass distribution for the
secondary. Fig. \ref{fig:chi2summary} summarizes the $\chi^2_{\nu,min}$ as a
function of the  binary fraction for five out of the six galaxies of the
LCID sample, based on the DAOPHOT photometry: Cetus, Tucana, LGS~3,
IC~1613, and Leo~A. The  five galaxies show a common general trend:  the
$\chi^2_\nu$ decreases for increasing number of binaries, and reaches a 
plateau and possibly a minimum around 40\%  -- 60\%. We conclude that we 
cannot put strong constraints on the binary content of these galaxies on 
the basis of these data, although a very low binary content seems to be
disfavored.

\begin{figure}
\epsscale{1.5}
%\plottwo{cum_binaries.eps}{binaries3.eps}
\plottwo{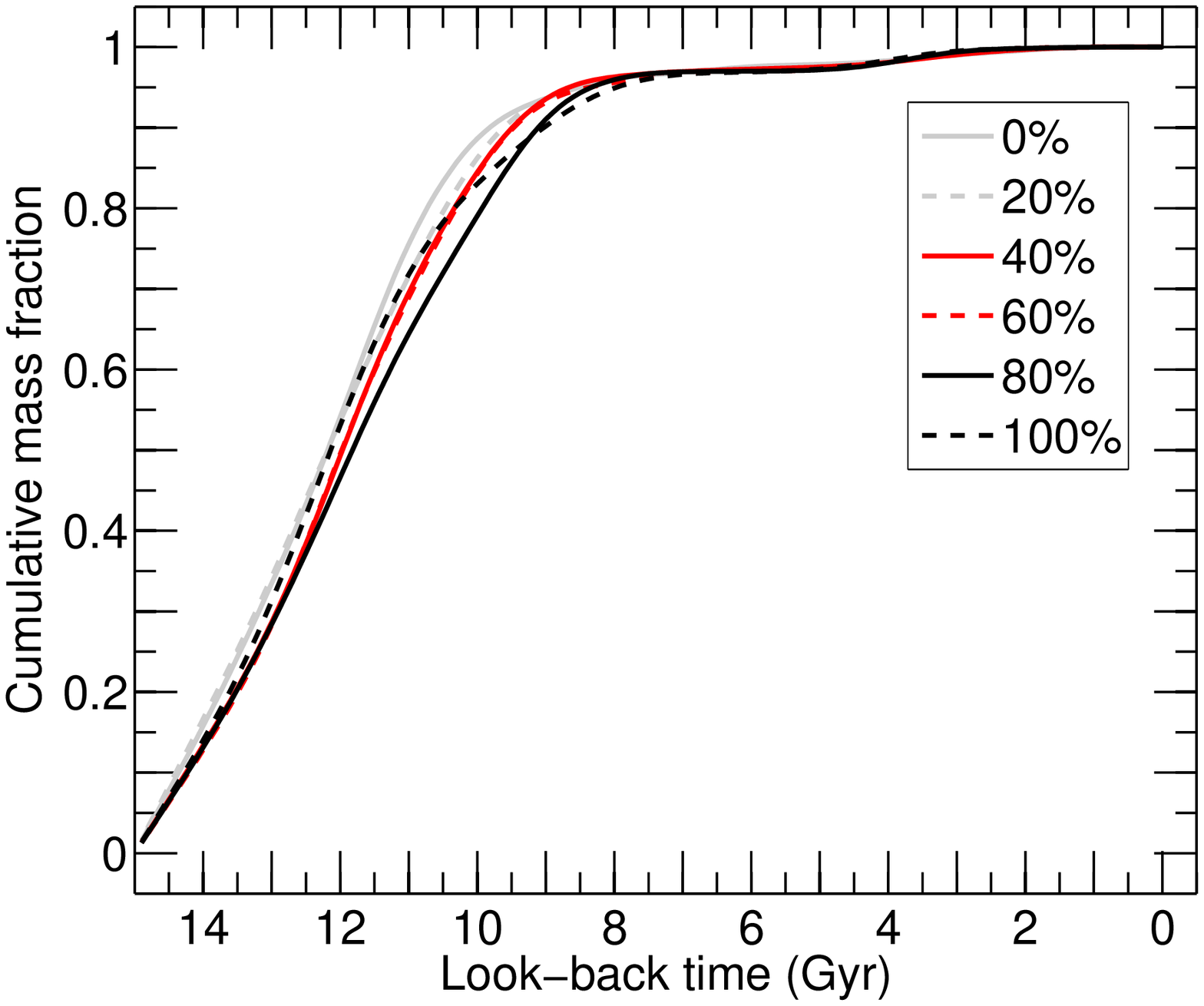}{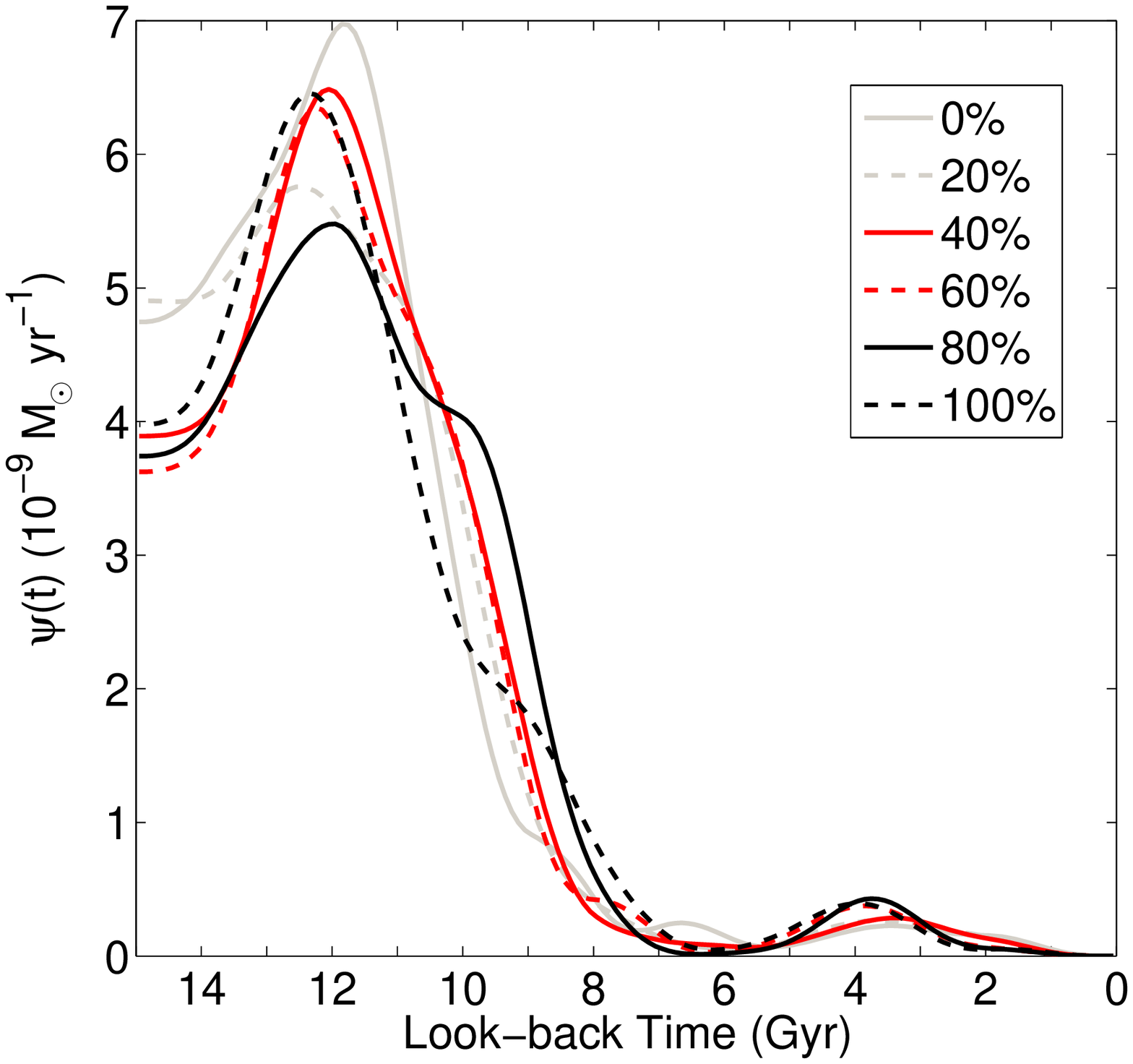}
\caption{{\itshape Top:} Cumulative mass fraction derived using model 
CMDs with six different percentages of binary stars. The amount of 
binaries slightly affects the solution, and we detect a mild tendency 
to recover slightly younger solutions for increasing amount of binaries.
{\itshape Bottom:} The six corresponding $\psi(t)$ show that all the 
main features do not vary for different assumptions of binaries. Note
that the main peak of star formation occur at the same age within $\sim$ 0.6 Gyr.
\label{fig:sfr4680}}
\end{figure}

The impact of different amount of binaries on $\psi(t)$ is shown in Fig.
\ref{fig:sfr4680},  where we plot the cumulative mass fraction and the
$\psi(t)$ for the six  different assumptions of the binary stars applied to
the DAOPHOT photometry of Cetus. We detect a mild trend, the solution
getting slightly younger for increasing amount of binaries. This is shown
also in the bottom panel of the same figure, where the $\psi(t)$ tends to
be more extended to younger ages for increasing binary fraction. However,
the main features are not strongly affected by the assumptions made on the
characteristics of the binary star population. In particular, the age of
the main peak of star formation is stable within $\sim$ 0.6 Gyr.

Moreover, we show that the {\itshape position} of the minimum in the 
$\chi^2_\nu$  plot as a function of $\delta_{col} - \delta_{mag}$, changes
as a function of the assumed binary content. As a  general trend, we
found that for increasing binary content, the minimum $\chi^2_\nu$ tends to
migrate toward redder color and shorter distance, as summarized in Tab.
\ref{tab:tab3} in the case of Cetus. This can be easily understood in terms
of  how the increasing number of binaries changes the distribution of stars
in the model CMD. The higher the binaries content, the redder and more
luminous the MS and MSTO.

%%%%%%%%%%%%%%%%%%%%%%%%%%%%%%%%%%%%%%%%%%%%%%%%%%%%%%%%%%%%%%%%%%%%%%%%%%%%%%%%
%%%%%%%%%%%%%%%%%%%%%%%%%%%%%%%%%%%%%%%%%%%%%%%%%%%%%%%%%%%%%%%%%%%%%%%%%%%%%%%%

\begin{deluxetable}{cccc}
\tabletypesize{\scriptsize}
\tablewidth{0pt}
\tablecaption{Integrated quantities derived for the Cetus dSph.\label{tab:tab3}}
\tablehead{
\colhead{Bin. \%} & \colhead{Best $\chi^2_\nu$} & \colhead{$\Delta col-\Delta mag$} & \colhead{$\chi^2_\nu$at (0;0)}}
\startdata
   0   &   2.60$\pm$0.03 &   (-0.020 ; +0.1125)      &  4.53  \\
  20   &   2.18$\pm$0.04 &   (-0.010 ; +0.075)	     &  3.04  \\
  40   &   1.95$\pm$0.02 &   ( 0.0   ; +0.075)	     &  2.54  \\
  60   &   1.86$\pm$0.05 &   (+0.020 ; +0.075)       &  2.10  \\
  80   &   1.80$\pm$0.06 &   (+0.010 ; +0.0375)      &  1.88  \\
 100   &   1.75$\pm$0.05 &   (+0.020 ; +0.0375)      &  1.84  \\

\enddata
%\tablenotetext{a}{}
\end{deluxetable}

%%%%%%%%%%%%%%%%%%%%%%%%%%%%%%%%%%%%%%%%%%%%%%%%%%%%%%%%%%%%%%%%%%%%%%%%%%%%%%%%
%%%%%%%%%%%%%%%%%%%%%%%%%%%%%%%%%%%%%%%%%%%%%%%%%%%%%%%%%%%%%%%%%%%%%%%%%%%%%%%%

%%%%%%%%%%%%%%%%%%%%%%%%%%%%%%%%%%%%%%%%%%%%%%%%%%%%%%%%%%%%%%%%%%%%%%
%%%%%%%%%%%%%%%%%%%%%%%%%%%%% FIG 22 %%%%%%%%%%%%%%%%%%%%%%%%%%%%%%%%%

%%%%%%%%%%%%%%%%%%%%%%%%%%%%%%%%%%%%%%%%%%%%%%%%%%%%%%%%%%%%%%%%%%%%%%
%%%%%%%%%%%%%%%%%%%%%%%%%%%%%%%%%%%%%%%%%%%%%%%%%%%%%%%%%%%%%%%%%%%%%%

%%%%%%%%%%%%%%%%%%%%%%%%%%%%%%%%%%%%%%%%%%%%%%%%%%%%%%%%%%%%%%%%%%%%%%
%%%%%%%%%%%%%%%%%%%%%%%%%%%%% FIG 23 %%%%%%%%%%%%%%%%%%%%%%%%%%%%%%%%%

%%%%%%%%%%%%%%%%%%%%%%%%%%%%%%%%%%%%%%%%%%%%%%%%%%%%%%%%%%%%%%%%%%%%%%
%%%%%%%%%%%%%%%%%%%%%%%%%%%%%%%%%%%%%%%%%%%%%%%%%%%%%%%%%%%%%%%%%%%%%%

\end{document}